\documentclass[aps,superscriptaddress,twocolumn,twoside,floatfix,pra,nofootinbib,a4paper, accepted=2021-12-31]{quantumarticle}

\pdfoutput=1

\usepackage{epsfig}
\usepackage{amsfonts}
\usepackage{amsmath}
\usepackage{amssymb}
\usepackage{amsthm}
\usepackage{IEEEtrantools}
\usepackage{color}
\usepackage{multirow}
\usepackage[normalem]{ulem}
\newcommand{\stkout}[1]{\ifmmode\text{\sout{\ensuremath{#1}}}\else\sout{#1}\fi}
\usepackage{latexsym}
\usepackage{mathrsfs}
\usepackage{natbib}
\usepackage{verbatim}
\usepackage[T1]{fontenc}
\usepackage{float}
\usepackage{dsfont}
\usepackage{mathtools}
\usepackage{bbold} % Turn this on!!!
\usepackage{graphicx}
\usepackage[colorlinks=true,linkcolor=blue,citecolor=magenta,urlcolor=blue]{hyperref}

\DeclareMathOperator{\Tr}{Tr}
\DeclareMathOperator{\conv}{Conv}
\DeclareMathOperator{\cone}{Cone}

\newcommand{\ket}[1]{\lvert #1 \rangle}
\newcommand{\bra}[1]{\langle #1 \rvert}
\newcommand{\braket}[2]{\langle #1 \vert #2 \rangle}

\newcommand{\ketbra}[2]{\lvert #1 \rangle \langle #2 \rvert}
\newcommand{\proj}[1]{\ketbra{#1}{#1}}

\newcommand{\abs}[1]{\lvert #1 \rvert}
\newcommand{\babs}[1]{\bigl\lvert #1 \bigr\rvert}
\newcommand{\vect}[1]{\boldsymbol{#1}}

\newcommand{\bigro}[1]{\bigl(#1\bigr)}
\newcommand{\Bigro}[1]{\Bigl(#1\Bigr)}

\newcommand{\bigsq}[1]{\bigl[#1\bigr]}

\newcommand{\bigbr}[1]{\bigl\{#1\bigr\}}

\newcommand{\bnorm}[1]{\bigl\lVert#1\bigr\rVert}
\newcommand{\btrnorm}[1]{\bnorm{#1}_{1}}

\newcommand{\cdet}{\mathcal{C}_{\text{det}}}

\newcommand{\qdet}{\mathcal{Q}_{\text{det}}}

\newcommand{\rmin}{\text{min}}
\newcommand{\guess}{\mathrm{g}}

\newcommand{\gap}{D\'epartement de Physique Appliqu\'ee, Universit\'e de Gen\`eve, CH-1211 Gen\`eve, Switzerland}

\newcommand{\liq}{Laboratoire d'Information Quantique, CP~225, Universit\'{e} libre de Bruxelles (ULB), \\
  Av.\ F.~D.~Roosevelt 50, 1050 Bruxelles, Belgium}

\begin{document}

%%%%%%%%%%%%%%%%%%%%%%%%%%%%%%%%%%%%%%%%%%%%%%%%%%%%%%%%%%%%%%%%%%%

\title{Informationally restricted correlations: a general framework for classical and quantum systems}

%%%%%%%%%%%%%%%%%%%%%%%%%%%%%%%%%%%%%%%%%%%%%%%%%%%%%%%%%%%%%%%%%%%

\author{Armin Tavakoli}
\affiliation{\gap}
\affiliation{Institute for Quantum Optics and Quantum Information -- IQOQI Vienna, Austrian Academy of Sciences, Boltzmanngasse 3, 1090 Vienna, Austria}

\author{Emmanuel Zambrini Cruzeiro }
\author{Erik Woodhead}
\author{Stefano Pironio}
\affiliation{\liq}
%\date{23~October~2020}
%%%%%%%%%%%%%%%%%%%%%%%%%%%%%%%%%%%%%%%%%%%%%%%%%%%%%%%%%%%%%%%%%%%
\begin{abstract}
We introduce new methods and tools to study and characterise classical and quantum correlations emerging from prepare-and-measure experiments with informationally restricted communication. We consider the most general kind of informationally restricted correlations, namely the ones formed when the sender is allowed to prepare statistical mixtures of mixed states, showing that contrary to what happens in Bell nonlocality, mixed states can outperform pure ones. We then leverage these tools to derive device-independent witnesses of the information content of quantum communication, witnesses for different quantum information resources, and demonstrate that these methods can be used to develop a new avenue for semi-device independent random number generators.
\end{abstract}

%%%%%%%%%%%%%%%%%%%%%%%%%%%%%%%%%%%%%%%%%%%%%%%%%%%%%%%%%%%%%%%%%%%

\maketitle

%%%%%%%%%%%%%%%%%%%%%%%%%%%%%%%%%%%%%%%%%%%%%%%%%%%%%%%%%%%%%%%%%%%

\section{Introduction}
\label{sec:introduction}

\begin{figure}[t!]
  \centering
  \includegraphics[width=\columnwidth]{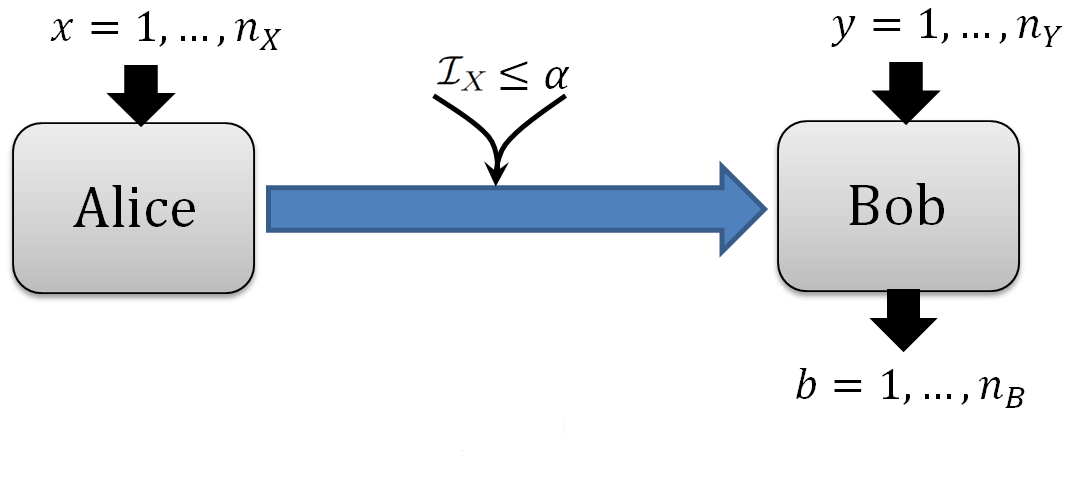}
  \caption{Illustration of prepare-and-measure experiment in which the communication is restricted to carry at most $\alpha$ bits of information about $X$}\label{FigScenario}
\end{figure}

Consider an experiment of the kind illustrated in Fig.~\ref{FigScenario},
where a sender, Alice, selects an input $x \in \{1, \dotsc, n_{X}\}$, encodes
it into some physical system and transmits it to a receiver, Bob. Bob
performs on the incoming system some measurement, represented by an input
$y \in \{1, \dotsc, n_{Y}\}$, and gets an outcome
$b\in \{1, \dotsc, n_{B}\}$. This \emph{prepare-and-measure} experiment is
ubiquitous in physics and forms the basis of many communication systems.

The transmission of physical messages between Alice and Bob serves to establish certain correlations between them. These correlations can be fully characterised by the set of probabilities $p(b|x,y)$ which represent how, for a given measurement $y$ performed by Bob, his outcome $b$ depends on Alice's input $x$. In full generality, we can associate to each input $x$ selected by Alice a quantum state $\rho_x$ and to each measurement $y$ selected by Bob a positive operator-valued measure (POVM) $\{M_{b|y}\}_b$, so that we can write
\begin{equation}
  \label{eq:born}
  p(b|x,y) = \Tr \bigsq{\rho_{x} M_{b|y}}.
\end{equation}
The special case where Alice and Bob are manipulating classical systems, instead of quantum ones, can be treated analogously by taking the states and measurements to be diagonal in the same basis:
\begin{IEEEeqnarray}{rCl}
  \label{eq:classical1}
  \rho_{x} &=& \sum_{m} p(m|x) \proj{m} , \\
  \label{eq:classical2}
  M_{b|y} &=& \sum_{m} p(b|y,m) \proj{m} ,
\end{IEEEeqnarray}
where the variable $m$ denotes the possible values of Alice's classical message. 

In this work, we are interested in characterising what kind of correlations between Alice and Bob, i.e., which set of probabilities $p(b|x,y)$, are possible under the sole restriction of some constraint on the communication capabilities of the classical or quantum systems $\rho_{x}$ emitted by Alice. 

To date, the most commonly considered communication constraint in this setting has been a bound on the Hilbert-space dimension $d$ of the emitted quantum systems (corresponding to the number of different possible messages $m$ in the classical case). In the last two decades, a large body of works has investigated the interplay between correlations and dimension in this setting \cite{Ambainis1998, Galvao2001, Trojek2005, Ambainis2008, Gallego2010, Brunner2013, Tavakoli2015, Navascues2015, Smania2016, Zukowski2017, Martinez2018}. This line of work led, e.g., to the notion of dimension witnesses \cite{Brunner2008,Gallego2010} and to semi-device-independent protocols \cite{Pawlowski2011}, such as randomness generation  \cite{Li2011}, quantum key distribution \cite{Woodhead2015}, and self-testing \cite{Kaniewski2018}. Evidently, a quantum or classical $d$-dimensional system can carry at most $\log_2 d$ bits of information and thus a bound on the dimension represents an information constraint. However, the physical dimension does not provide a complete picture of the concept of information. For instance, there are many systems of dimension $d'>d$ that do not carry more than $\log_{2} d$ bits of information. Furthermore, in practical semi-device-independent protocols, assuming an exact bound on the dimension may be problematic to justify (a fact that has partly motivated other recent approaches \cite{VanHimbeeck2017, Brask2017, Entropy, Distrust}). A more satisfying and practically relevant approach may be to constrain the communication in terms of a continuous information measure. 

Following \cite{Tavakoli2020}, we specify here the communication constraint
on the physical systems $\rho_x$ received by Bob as an upper bound
\begin{equation}\label{eq:pguess-bound}
P_{\guess}(X|B)\leq G
\end{equation}
on the guessing probability of the input $X$\footnote{$X$ and $B$ are random variables.},
\begin{equation}
  P_{\guess}(X|B)
  = \max_{\{N_{x}\}_x} \sum_{x} q_{x} \Tr \bigsq{\rho_{x} N_{x}} ,
\end{equation}
where the maximisation is taken over all possible POVMs $\{N_{x}\}_x$ on the physical system that Bob receives. This guessing probability represents the optimum average probability with which Bob would correctly guess Alice's input $x$ if he were to perform an ideal POVM on the incoming messages $\rho_x$, assuming that Alice selects each input with prior probability  $q_x$ \cite{Konig2009}. The guessing probability $P_{\guess}(X|B)$ can take any value from $P_{\guess}(X|B)=\max_{x} q_{x}$ when the states are the same and hence carry no information about Alice's input (in which case Bob's best guessing strategy is to output the most probable input $x$ according to $q_x$), up to $P_{\guess}(X|B)=1$ when they are perfectly distinguishable.   Different communication restrictions on the messages can be specified by the choice of the bound $G$, as well as the input probabilities $q_x$. 

Equivalently, one can express the communication restriction \eqref{eq:pguess-bound} as an upper bound $\mathcal{I}(X|B) \leq \alpha$ on the information measure 
\begin{equation}\label{eq:infodef}
  \mathcal{I}(X|B) = H_{\rmin}(X) - H_{\rmin}(X|B),
\end{equation}
defined in term of the min-entropies $H_{\rmin}(X) = -\log_{2}
\bigro{\max_{x} \{q_{x}\}}$ and $H_{\rmin}(X|B) = -\log_{2}
\bigro{P_{\guess}(X|B)}$. This quantity, expressed in bits, ranges from
$\mathcal{I}(X|B) = 0$ when the states carry no information about Alice's
input, up to $\mathcal{I}(X|B) = \log_{2}(n_{X})$ bits, when they are
perfectly distinguishable and chosen equiprobably, i.e., $q_{x} = 1/n_{X}$. There exist in principle a number of different other information measures that we could consider (see e.g.~\cite{Ciganovic2013}) but the one we choose has a clear operational meaning and is convenient to work with.

We emphasise that $q_x$ does not represent the actual prior from which Alice selects her input. Instead, it is a part of the assumption on Alice's source. Indeed, we are interested here in constraining \emph{conditional} probabilities $p(b|x,y)$ which therefore do not depend on any prior probabilities with which Alice's input $x$ and Bob's inputs $y$ are selected. To constrain these conditional probabilities $p(b|x,y)$ we make a certain assumption about the source, specifically about the information-capacity of the ensemble of states $\{\rho_x\}$ it prepares. This information-capacity can be defined in various ways. The definition  we chose here can be thought of as a fictious game: how well the classical variable $x$ could correctly be identified by Bob if it were encoded by Alice in the state $\rho_x$ and chosen with probability $q_x$. In the same way that the optimal measurement performed to guess $x$ in this fictious game is not necessarily the same as the actual measurements taking place in Bob's measurement apparatuses and leading to the conditional probabilities $p(b|x,y)$, the prior probabilities $q_x$ need not be the same as the prior probabilities $p_x$ used by Alice to select her input in any actual scenario or protocol involving the conditional probabilities $p(b|x,y)$. In particular, a given scenario, say a DIRNG protocol where Alice's select her input with some fixed probabilities $p_x$, can be analyzed using different choices of $q_x$, this simply correspond to different assumptions about the source.

Note that one can also completely eliminate $q_x$ from the analysis by choosing the uniform prior $q_x=1/n_X$ (where $n_X$ denotes the number of inputs of Alice). For a bound of the form $\mathcal{I}(X|B) \leq \alpha$, this corresponds to the strongest assumption on the source in the sense that  $\mathcal{I}(X|B)_\text{uni} \leq \alpha$ implies $\mathcal{I}(X|B)_\text{bias} \leq \alpha$ for any choice of biased distribution $q_x$, as shown in \cite{NewPaper}.

Finally, we remark that instead of viewing the bound \eqref{eq:pguess-bound} as characterizing the preparations of Alice, we can alternatively view it as a constraint on the channel relating Alice to Bob. Indeed, $\epsilon=1-G$ can be understood as an upper-bound on the average\footnote{The reference \cite{Renner2012} defines the error $\epsilon$ for uniform prior, but one can also generalize this concept for abitrary priors $q_x$.} error through which a classical message of size $n_X$ can be communicated in one shot through the channel for whatever encoding Alice may choose \cite{Renner2012}.

We develop here a versatile toolbox for characterising the set of probabilities $p(b|x,y)$ that are possible given arbitrary information constraints $P_{\guess}(X|B)\leq G$ (or, equivalently, $\mathcal{I}(X|B) \leq \alpha$). Our approach is fully general and does not make any assumptions about the states and measurements beyond the information constraint, and in particular no assumptions about their dimension. In the classical case, we provide a characterisation of the set of informationally restricted correlations in terms of linear programming and in the quantum case through a hierarchy of semidefinite programming relaxations. We also show, in analogy with the dimension bounded case, how to apply our methods to construct device-independent witnesses of communication (quantified in terms of our information measure), resource inequalities for classical and quantum systems carrying one bit of information, and semi-device-independent random number generation (RNG) protocols. In particular, we will show concrete examples of high-rate RNG and also demonstrate that data obtained in RNG experiments assuming a 1-qubit bound can be recycled to certify the same amount of randomness under the strictly weaker assumption of a 1-bit information bound.

Our work can be seen as a follow-up to Ref.~\cite{Tavakoli2020}, which originally proposed to replace the dimension bound in semi-device-independent scenarios by the information bound $P_{\guess}(X|B)\leq G$ (or $\mathcal{I}(X|B) \leq \alpha$) considered here. However, \cite{Tavakoli2020} implicitly modelled the correlations established between Alice and Bob as statistical mixtures
\begin{equation}\label{eq:convexsum}
p(b|x,y) =  \sum_{\lambda} p(\lambda) p_\lambda(b|x,y)
\end{equation}
of correlations $p_\lambda(b|x,y)$ obtained by measuring \emph{pure} states:
\begin{equation}\label{eq:pure}
  p_\lambda(b|x,y) = \bra{\psi_x^{(\lambda)}} M_{b|y}^{(\lambda)} \ket{\psi_x^{(\lambda)}} .
\end{equation}
The guessing probability constraining the communication was then defined as the following averaged quantity over the classical shared variable $\lambda$:
\begin{equation}\label{eq:guesspure}
  P_{\guess}(X|B) 
  = \sum_\lambda p(\lambda) \max_{\{N^{(\lambda)}_{x}\}_x} \sum_{x} q_{x} \bra{\psi_x^{(\lambda)}} N_{x}^{(\lambda)} \ket{\psi_x^{(\lambda)}} .
\end{equation}
Similarly, in the classical case, the correlations between Alice and Bob were modelled as statistical mixtures of correlations 
\begin{equation}\label{eq:lambdaclass}
  p_\lambda(b|x,y) = \sum_m \delta(m,m_x^{(\lambda)})\, p_\lambda(b|y,m) \,,
\end{equation}
established by sending \emph{deterministic} messages $m_x^{(\lambda)}$ for given $x$ and $\lambda$. 

The sets of such pure state correlations, in the quantum case, or deterministic correlations, in the classical case, compatible with a given communication constraint $P_{\guess}(X|B) \leq G$ are easily seen to be particular subcases of the more general correlations that we consider here. Indeed, they can be obtained by assuming the states and measurements in \eqref{eq:born} to take the following specific forms
\begin{IEEEeqnarray}{rCl}
  \rho_{x} &=& \sum_\lambda p(\lambda)\, \proj{\lambda}\otimes \proj{\psi^{(\lambda)}_x} , \label{eq:conv1}\\
  M_{b|y} &=& \sum_\lambda \proj{\lambda}\otimes M_{b|y}^{(\lambda)} \label{eq:conv2}
\end{IEEEeqnarray}
in the quantum case, and
\begin{IEEEeqnarray}{rCl}
  \rho_{x} &=& \sum_{\lambda,m} p(\lambda) p_\lambda(m|x)\,  \proj{\lambda}\otimes\proj{m} , \\
  M_{b|y} &=& \sum_{\lambda,m} p(\lambda) p_\lambda(b|y,m)\, \proj{\lambda}\otimes\proj{m}
\end{IEEEeqnarray}
in the classical case, which recovers both the convex sum \eqref{eq:convexsum} and the average guessing probability \eqref{eq:guesspure}. 
Interestingly, while in more traditional works on correlations, such as in the study of Bell nonlocality \cite{Brunner2014}, statistical mixtures of pure states (or of deterministic correlations) generate the full set of correlations, they only represent a proper subset of the possible correlations in our information-restricted setting. This is because given a set of arbitrary states $\rho_x$ satisfying the information constraint $P_{\guess}(X|B) \leq G$, one can generally not re-interpret them as a mixtures of pure states without increasing their distinguishability, hence potentially violating the condition $P_{\guess}(X|B) \leq G$. 

The formulation we consider here is fully general and does not make any implicit assumption on the structures of the states appearing in the definition \eqref{eq:born}. Throughout the paper, we will compare our results to those that would be obtained under the pure-state approach of \cite{Tavakoli2020} in order to illustrate the differences in the two formulations.

%%%%%%%%%%%%%%%%%%%%%%%%%%%%%%%%%
%%%%%%%%%%%%%%%%%%%%%%%%%%%%%%%%%	
%%%%%%%%%%%%%%%%%%%%%%%%%%%%%%%%%
%%%%%%%%%%%%%%%%%%%%%%%%%%%%%%%%%

\section{Basic properties and simple scenarios}
\label{sec:GboundMessages}

In the following, we refer to the prepare-and-measure scenario of Fig.~\ref{FigScenario}, with $n_X$ inputs for Alice, $n_Y$ inputs for Bob, and $n_B$ outputs, as a $(n_{X}, n_{Y}, n_{B})$-scenario. Given an information bound specified by a probability distribution $q_x$ and a number $G\in[\max_x\{q_x\},1]$, we denote by $\mathcal{Q}$ the set of quantum correlations compatible with that information bound, i.e., the set of probability distributions $p(b|x,y)$ for which there exist states $\rho_x$ and measurement operators $M_{b|y}$ defined on some Hilbert space of arbitrary dimension $d$ that satisfy the Born rule \eqref{eq:born} and the constraint \eqref{eq:pguess-bound}. Similarly, $\mathcal{C}$ denotes the set of classical correlations, i.e., satisfying in addition \eqref{eq:classical1}--\eqref{eq:classical2}. By plugging this specific form for the states and measurements in \eqref{eq:born} and \eqref{eq:pguess-bound}, classical correlations can also be defined as those that can be written as
\begin{equation}\label{eq:class}
p(b|x,y)=\sum_m p(m|x)p(b|y,m)
\end{equation}
and satisfying the information constraint
\begin{equation}\label{eq:information-bound}
P_{\guess}(X|B)
  = \sum_{m} \max_x q_{x} p(m|x) \leq G ,
\end{equation}
since the optimal POVM $\{N_x\}$ in this case is the one that reads the classical message $m$ and outputs the value $x$ that maximises $q_x p(m|x)$.

The sets $\mathcal{Q}$ and $\mathcal{C}$ are easily seen to be convex, using a construction akin to \eqref{eq:conv1} and \eqref{eq:conv2}. That is, we can without loss of generality assume that the states sent by Alice and the measurements performed by Bob depend on some shared randomness $\lambda$ (independent of $x$).

As a consequence, when writing the correlations explicitly as a convex sum \eqref{eq:convexsum}, we can without loss of generality assume Bob's measurements to be extremal conditioned on $\lambda$: if the measurements of Bob depend on some local randomness, we can always incorporate it instead in the shared randomness $\lambda$. In the classical case $\mathcal{C}$, this means that we can without loss of generality assume Bob's classical response $p_\lambda(b|y,m)$ to be deterministic, i.e., such that $p_\lambda(b|y,m)\in\{0,1\}$. 
However, as noted earlier, we cannot without loss of generality assume the states to be pure (or deterministic in the classical case) when conditioned on $\lambda$ as rewriting a mixed-state as a convex combination of pure states could violate the original guessing probability bound. 

The sets $\mathcal{Q}$ and $\mathcal{C}$ satisfy certain basic inequalities.
Obviously, since the $p(b|x, y)$ are probabilities, they must by definition satisfy the positivity and normalisation conditions
\begin{IEEEeqnarray}{c+c}
  \label{eq:positivity}
  p(b|x, y) \geq 0 , & \forall b, x, y
\end{IEEEeqnarray}
and
\begin{IEEEeqnarray}{c+c}
  \label{eq:normalisation}
  \sum_{b} p(b|x, y) = 1 , & \forall x, y .
\end{IEEEeqnarray}
In addition, since post-processing cannot improve the distinguishability
between messages and since all measurements $\{M_{b|y}\}_b$ of Bob can be
viewed as (typically suboptimal) information-extraction POVMs, it holds that
\begin{IEEEeqnarray}{c+c}
  \label{eq:pbxy-pguess-bound}
  \sum_{b} \max_{x} q_{x} p(b|x, y) \leq G , & \forall y ,
\end{IEEEeqnarray}
since, as in \eqref{eq:information-bound}, when Bob gets the result $b$ when
he performs the measurement corresponding to input $y$, his best guess of $x$
is the value that maximises $q_{x} p(b|x,y)$. This last constraint can
explicitly be rewritten as a series of linear inequalities
\begin{IEEEeqnarray}{c+l}
  \label{eq:max-to-linear-ineqs}
  \sum_{b} q_{x_{b}} p(b|x_{b}, y) \leq G ,
  & \forall y,\, \forall \vect{x}=(x_1,x_2,\dotsc,x_{n_B})
  \IEEEeqnarraynumspace
\end{IEEEeqnarray}
where $\vect{x} = (x_1,x_2,\dotsc,x_{n_B})\in\{1,\dotsc,n_X\}^{\times n_B}$
assigns to each output $b$ a value $x_b$.

We remark that, though it is harmless to specify them, not all of the
inequalities \eqref{eq:max-to-linear-ineqs} are always relevant as they may
already be implied by normalisation and positivity of the probabilities alone
(as well potentially as constraints specific to $\mathcal{C}$ and
$\mathcal{Q}$). Precisely which ones are redundant depends on the upper bound
$G$ chosen. The instances with all the components of $\vect{x}$ equal
($x_{1} = x_{2} = \dotsb = x_{n_{B}}$) in particular are always redundant as
the left side of \eqref{eq:max-to-linear-ineqs} is in these cases always
upper bounded by the smallest possible value, $\max_{x} \{q_{x}\}$, of the
guessing probability. At the opposite extreme, \eqref{eq:pbxy-pguess-bound}
always becomes redundant entirely for sufficiently high $G$ when Alice's
device has more inputs than Bob's has outcomes. This, supposing we label
Alice's inputs so that $q_{1} \geq q_{2} \geq \dotsb \geq q_{n_{X}}$, is
because the left side of \eqref{eq:pbxy-pguess-bound} is also always bounded
by
\begin{equation}
  \sum_{b} \max_{x} q_{x} p(b|x, y) \leq \sum_{x=1}^{n_{B}} q_{x} ,
\end{equation}
which is strictly less than one if Alice has more than $n_{B}$ inputs that are
used with nonzero probability.

The set of correlations satisfying Eqs.~\eqref{eq:positivity},
\eqref{eq:normalisation}, and \eqref{eq:pbxy-pguess-bound} is a polytope
$\mathcal{G}$. The polytope $\mathcal{G}$ can be interpreted as the set of
correlations attainable under informational restrictions when no assumption
is made on the underlying physical theory. Therefore, recalling also that the
classical set is contained in the quantum set, we have the inclusions
$\mathcal{C}\subseteq \mathcal{Q}\subseteq \mathcal{G}$.

An important first step in semi-device-independent approaches is to establish that one can distinguish between classical and quantum correlations, i.e., that $\mathcal{C}\subset \mathcal{Q}$. We show here below that in the simplest case of communication experiments with only two inputs on Alice ($n_X=2$), the classical, quantum and theory-independent sets are identical ($\mathcal{C}=\mathcal{Q}=\mathcal{G}$). Notably, this stands in contrast to other established approaches to semi-device-independence \cite{Brask2017, VanHimbeeck2017}. Later, we will find that $\mathcal{C}\subset \mathcal{Q}$ indeed is possible when Alice has more than two inputs. In sections~\ref{sec:classical} and \ref{sec:tools} we describe how to characterise the classical set and quantum set, respectively, in a general and systematic manner.

\subsection{$\mathcal{C}=\mathcal{Q}=\mathcal{G}$ when Alice has $n_X=2$ inputs}
\label{sec:stochastic-same}

We show that for $n_X=2$ it holds that $\mathcal{C}=\mathcal{Q}=\mathcal{G}$ by proving that every $p(b|x,y)\in\mathcal{G}$ admits a classical model. To this end, note that the constraints Eqs.~\eqref{eq:positivity}--\eqref{eq:pbxy-pguess-bound} are decoupled with respect to $y$.  In other words, for each individual value of $y$, we obtain a separate polytope and the full set of probabilities is just the Cartesian product of the $n_{Y}$ identical polytopes corresponding to the individual values of $y$. We derive the vertices of these polytopes in Appendix~\ref{sec:vertex-probabilities}. For $n_{B} = 3$ (which is representative), up to permutations of Bob's outputs they are
\begin{IEEEeqnarray}{rCl}
  \label{eq:vertex-v1}
  \vect{v}_{1}(y) &=& \begin{pmatrix}
    1 & 0 & 0 \\
    1 & 0 & 0
  \end{pmatrix} , \\
  \vect{v}_{2}(y) &=& \begin{pmatrix}
    1 & 0 & 0\\
    \frac{1 - G}{q_2} & 1 - \frac{1 - G}{q_2} & 0
  \end{pmatrix} , \\
  \vect{v}_{3}(y) &=& \begin{pmatrix}
    \frac{1 - G}{q_1} & 1 - \frac{1 - G}{q_1} & 0 \\
    1 & 0 & 0
  \end{pmatrix} , \\
  \label{eq:vertex-v4}
  \vect{v}_{4}(y) &=& \begin{pmatrix}
    \frac{1 - G}{q_1} & 1 - \frac{1 - G}{q_1} & 0 \\
    \frac{1 - G}{q_2} & 0 & 1 - \frac{1 - G}{q_2}
  \end{pmatrix} ,
\end{IEEEeqnarray}
where we use a matrix notation
\begin{equation}
  \vect{v}_{j}(y) = \begin{pmatrix}
    p(1|1, y) & p(2|1, y) & \cdots \\
    p(1|2, y) & p(2|2, y) & \cdots
  \end{pmatrix} 
\end{equation}
to summarise the probabilities $p(b|x,y)$ defining each vertex $\vect{v}_j$. 
The vertices for $n_{B} \neq 3$ are trivial variations of those above: for $n_{B} > 3$ the vertices are the same except with additional columns of zeros while for $n_{B} < 3$ we simply discard the vertices that have more than $n_{B}$ columns with nonzero entries in them. 

Crucially, all the vertices $\vect{v}_{1}(y)$--$\vect{v}_{4}(y)$, including all their permutations, can be generated by performing different measurements on the same two commuting (classical) states
\begin{IEEEeqnarray}{rCl}
  \label{eq:rho1-diagonal}
  \rho_{1} &=& \frac{1-G}{q_{1}} \proj{0}
  + \Bigro{1 - \frac{1-G}{q_{1}}} \proj{1} , \\
  \label{eq:rho2-diagonal}
  \rho_{2} &=& \frac{1-G}{q_{2}} \proj{0}
  + \Bigro{1  - \frac{1 - G}{q_{2}}} \proj{2} .
\end{IEEEeqnarray}
For example, the vertex $\vect{v}_{3}(y)$ is obtained by measuring $\{M_{b|y}\}$ with
\begin{IEEEeqnarray}{rCl}
  M_{1|y} &=& \proj{0} + \proj{2} , \\
  M_{2|y} &=& \proj{1} , \\
  M_{3|y} &=& 0 .
\end{IEEEeqnarray}
Furthermore, any convex mixtures of vertices of the kind above, which is to say, any probability $p(b|x,y)$ satisfying the conditions \eqref{eq:positivity}--\eqref{eq:pbxy-pguess-bound} above, can be generated by performing the corresponding convex mixtures of POVMs on Bob's side. We conclude that Eqs.~\eqref{eq:positivity}, \eqref{eq:normalisation}, and \eqref{eq:pbxy-pguess-bound} completely characterise both $\mathcal{C}$, $\mathcal{Q}$ and $\mathcal{G}$.

\subsection{Inequivalence of general correlations and pure-state correlations}
\label{sec:inequivalence}

Following \cite{Tavakoli2020}, we denote by $\mathcal{Q}_{\text{pure}}\subseteq \mathcal{Q}$ the subset of $\mathcal{Q}$ consisting of convex combination of pure-state correlations \eqref{eq:pure} and $\mathcal{C}_{\text{det}}\subseteq \mathcal{C}$ the subset of $\mathcal{C}$ consisting of convex combinations of deterministic classical correlations  \eqref{eq:lambdaclass}. As we show below, already for the simplest communication scenario ($n_X=2$), we can distinguish between $\mathcal{Q}$ and $\mathcal{Q}_{\text{pure}}$ as well as between $\mathcal{C}$ and $\mathcal{C}_{\text{det}}$, i.e., $\mathcal{Q}_{\text{pure}}\subset \mathcal{Q}$ and  $\mathcal{C}_{\text{det}}\subset \mathcal{C}$.
Note, though, that the relation between $\mathcal{Q}_{\text{pure}}$ and $\mathcal{C}$ is more complex. We will see that in the simple scenario below that $\mathcal{Q}_{\text{pure}}\subset \mathcal{C}$. But in other scenarios one can have correlations in $\mathcal{Q}_{\text{pure}}$ that are outside $\mathcal{C}$ so that the two sets intersect, but none is strictly contained in the other. This justifies looking at the larger quantum set $\mathcal{Q}$, which by definition always satisfies $\mathcal{C}\subseteq \mathcal{Q}$ and thus can never be outperformed using classical correlations.

Before looking at the general $n_X=2$ case, let us first consider the exceptional situation that Alice's inputs are equiprobable ($q_1 = q_2 = 1/2$). The states \eqref{eq:rho1-diagonal} and \eqref{eq:rho2-diagonal} become
\begin{IEEEeqnarray}{rCl}
  \label{eq:rho1-diagonal-equiprobable}
  \rho_{1} &=& 2 (1-G) \proj{0} + (2 G - 1) \proj{1} , \\
  \label{eq:rho2-diagonal-equiprobable}
  \rho_{2} &=& 2 (1-G) \proj{0} + (2 G - 1) \proj{2} .
\end{IEEEeqnarray}
Consider now a deterministic classical strategy with one bit of shared randomness. Specifically, Alice receives either $\lambda = 1$ with probability $p(1) = 2(1 - G)$ or $\lambda = 2$ with probability $p(2) = 2 G - 1$. If $\lambda = 1$ Alice prepares the state $\proj{0}$, while if $\lambda = 2$ Alice prepares the state $\proj{x}$ depending on her input $x \in \{1, 2\}$. This strategy generates the same states as \eqref{eq:rho1-diagonal-equiprobable} and \eqref{eq:rho2-diagonal-equiprobable} on average and the average guessing probability is still $G$. Thus, all correlations in $\mathcal{G}$ can be obtained and one finds no difference between the various sets: $\mathcal{C}_{\text{det}}=\mathcal{C}=\mathcal{Q}_{\text{pure}}=\mathcal{Q}=\mathcal{G}$.

\begin{figure}
  \centering
  \includegraphics{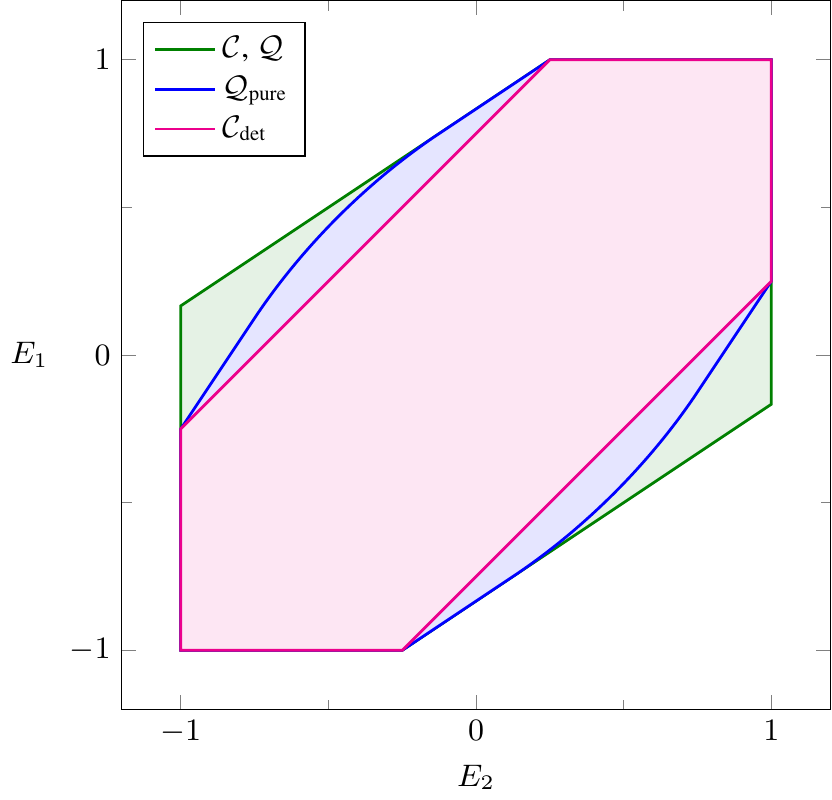}
  \caption{Informationally restricted correlations in the $(2,1,2)$ scenario with prior probabilities $(q_{1}, q_{2}) = (0.6, 0.4)$ and a bound $G = 3/4$ on the guessing probability (corresponding to $\mathcal{I}(X|B) \leq \log_{2}(5) - 2$ bits of information). The possible correlations are illustrated for deterministic classical strategies (magenta), deterministic quantum strategies (blue) and for classical and quantum stochastic strategies (green), which are the same in this case. \label{fig:212scenario}}
\end{figure}

In contrast, whenever the prior is biased ($q_1\neq q_2$), we find that the pure-state correlations and the general correlations are inequivalent (see Fig.~\ref{fig:212scenario}). Considering the scenario $(n_{X}, n_{Y}, n_{B}) = (2, 1, 2)$, the correlations can be characterised in terms of the expectation values
\begin{equation}
E_{x} = p(1|x) - p(2|x) 
\end{equation}
for $x = 1, 2$, where we have omitted $y$ due to its fixed value. In Appendix \ref{sec:212scenario-supp}, we show that the nontrivial facets of $\mathcal{C}$ and $\mathcal{Q}$ are 
\begin{equation}
\label{eq:cstoch-212-facets}
\abs{q_{1} E_{1} - q_{2} E_{2} } \leq 2 G - 1,
\end{equation}
in terms of the guessing probability bound $G$. The facets of $\mathcal{C}_{\text{det}}$ are likewise straightforward to derive due to the small number of possible deterministic strategies. We do this in Appendix~\ref{sec:212scenario-supp} and find that the nontrivial facets are 
\begin{equation}
\abs{E_{1} - E_{2}} \leq 2 \frac{G - q_{\text{max}}}{q_{\text{min}}} ,
\end{equation}
where $q_{\text{max}} = \max(q_{1}, q_{2})$ and $q_{\text{min}} = \min(q_{1}, q_{2})$. Whenever $q_{1} \neq q_{2}$ this bounds a strictly smaller set than \eqref{eq:cstoch-212-facets}.

Finally, we derive the exact boundaries of the set $\mathcal{Q}_{\text{pure}}$ in Appendix~\ref{sec:212scenario-supp}. Unlike the classical sets and $\mathcal{Q}$, this set is not a polytope. Aside from the trivial constraints $\abs{E_{x}} \leq 1$, it is bounded by an infinite family,
\begin{equation}
  \label{eq:qdet-212-boundary}
  \abs{c_{1} E_{1} - c_{2} E_{2}}
  \leq \sqrt{1 - \frac{4 c_{1} c_{2}}{q_{1} q_{2}} G (1 - G)} ,
\end{equation}
of linear inequalities, for parameters $c_{1}$ and $c_{2}$ satisfying $c_{1} + c_{2} = 1$ in the range $q_{\text{min}} \leq c_{1}, c_{2} \leq q_{\text{max}}$. This set is larger than $\mathcal{C}_{\text{det}}$ but smaller than $\mathcal{C}$ and $\mathcal{Q}$. Note that at the extreme $c_{1} = q_{1}$, \eqref{eq:qdet-212-boundary} reduces to \eqref{eq:cstoch-212-facets}. Hence, two flat parts of the boundary of $\mathcal{Q}_{\text{pure}}$ (see Fig.~\ref{fig:212scenario}) coincide with the nontrivial facets of $\mathcal{Q}$.

\section{Characterising classical correlations}
\label{sec:classical}

In this section, we explain how one can systematically determine the
boundaries of the classical set $\mathcal{C}$, which is a polytope; the
characterisation of the deterministic set $\mathcal{C}_{\text{det}}$ was
already addressed in \cite{Tavakoli2020}. We then apply our method to
explicitly derive the boundaries of $\mathcal{C}$ in the $(3,2,2)$ scenario
assuming Alice's inputs are chosen equiprobably, finding that $\mathcal{C}$
is strictly larger than $\mathcal{C}_{\text{det}}$ in this case. This differs
from the case with two inputs considered earlier, where $\mathcal{C}$ and
$\mathcal{C}_{\text{det}}$ were only found to be different when Alice's
inputs are not equiprobable. Finally we also point out how one can,
alternatively, generally test by linear
programming whether a correlation is in $\mathcal{C}$ or
not without explicitly needing to determine its boundaries.

\subsection{Identifying the boundaries of $\mathcal{C}$}

\subsubsection{General method}

The classical set $\mathcal{C}$ is, as mentioned above and as we have seen
explicitly for $n_{X} = 2$ in the previous section, a polytope and we could
in principle characterise it by determining its facets for any given upper
bound $G$ on the guessing probability. This direct approach would, however,
require us to rederive the facets of $\mathcal{C}$ for each value of $G$ that
we may be interested in. To avoid this we instead consider a related but
different set, which we call $\mathcal{C}^{+}$, of possible pairs
$\bigro{p(b | x, y), G}$ of probability distributions $p(b | x, y)$ and
guessing probability bounds $G$ that are compatible with \eqref{eq:class} and
\eqref{eq:information-bound}, which we repeat here for convenience:
\begin{IEEEeqnarray}{rCl}
  \label{eq:class2}
  p(b|x,y) &=& \sum_{m} p(m|x) p(b|y,m) , \\
  \label{eq:information-bound2}
  G &\geq& \sum_{m} \max_{x} q_{x} p(m|x) .
\end{IEEEeqnarray}
Casting the problem in this way allows us to derive the boundaries of the
classical set while leaving $G$ as a free variable.

The set $\mathcal{C}^{+}$ is clearly convex, as it is easily seen that
$\bigro{q\, p_1(b|x,y)+(1-q)\, p_2(b|x,y),\; q\,G_1+(1-q)\,G_2}$ belongs to
it if $(p_1(b|x,y),G_1)$ and $(p_2(b|x,y),G_2)$ do. To characterise it, it is
thus sufficient to characterise its extreme points and take their convex
hull.

As explained at the beginning of Section~\ref{sec:GboundMessages}, remember
that the extremal points of $\mathcal{C}$ have deterministic response
probabilities for Bob: $p(b|y,m)\in\{0,1\}$. If we fix such a deterministic
response for Bob, the probabilities $p(b|x,y)$ are then entirely determined
by the probability distribution $p(m|x)$ of Alice's messages. Those are
simply constrained by
\begin{IEEEeqnarray}{rCl}
  G-\sum_{m} \max_x q_{x} p(m|x) &\geq &0 , \\
  p(m|x) &\geq& 0 , \\
  \sum_{m} p(m|x) &=& 1 ,
\end{IEEEeqnarray}
which represents a finite set of linear inequalities for the set
$\mathcal{M}^{+}$ of possible pairs $\bigro{p(m|x), G}$ of message
probabilities and guessing probability bounds. The set $\mathcal{M}^{+}$ is
thus a polyhedron, i.e., an object like a polytope except that it is not
necessarily bounded\footnote{The set is not closed because the number $G$
  that we impose as an upper bound on the guessing probability is in
  principle unbounded. We could, of course, simply choose to impose a bound
  on it, such as $G \leq 1$. In that case, the set would become a (closed)
  polytope.}. Explicitly, this is a set $\mathcal{P} = \{\vect{p}\}$ of
points that can be generated from a finite number of vertices
$\vect{v}_{i} \in \mathcal{V}$ and conic generators
$\vect{w}_{j} \in \mathcal{W}$, i.e,
\begin{equation}
  \mathcal{P} = \conv(\mathcal{V}) + \cone(\mathcal{W})
\end{equation}
or, more explicitly, the set of points $\{\vect{p}\}$ that can be expressed
as
\begin{IEEEeqnarray}{rCl}
  \vect{p}
  &=& \sum_{i} \lambda_{i} \vect{v}_{i} + \sum_{j} \mu_{j} \vect{w}_{j} \\
&&\text{with } \lambda_i,\mu_j\geq 0, \quad \sum_{i} \lambda_{i} = 1 .
\end{IEEEeqnarray}

Provided that the number of possible messages $m$ is limited to a finite
number, the vertices and conic generators of $\mathcal{M}^+$ can be
determined using software such as PORTA or PANDA \cite{Lorwald2015}. In
Appendix~\ref{sec:finite-message-dim} we prove that every pair
$\bigro{p_\lambda (b|x,y), G_\lambda}$ can be constructed with a message
of size $2^{n_{X} - 1}$ without loss of generality. In practice, however, the
number of necessary messages may be considerably less than this in general:
in cases with two or three inputs where we explicitly determined the vertices
we never found that the number of necessary different messages exceeded the
number of inputs $n_{X}$.

Once the vertices and conic generators $\bigro{p(m|x), G}$ of
$\mathcal{M}^{+}$ have been obtained, one can generate all extreme points
$\bigro{p(b|x,y), G}$ of $\mathcal{C}^+$ using \eqref{eq:class2} for each of
the finite number of possible deterministic distributions $p(b|y,m)$ for
Bob. We thus find that $\mathcal{C}^+$ is described by a finite number of
vertices and conic generators, i.e., it is a polyhedron. Solving the facet
enumeration problem, which again can be done in software provided that the
problem is not too large, yields a finite number of inequalities that
completely characterises the set of points $\bigro{p(b|x,y), G}$ compatible
with classical stochastic communication.

\subsubsection{Boundaries of $\mathcal{C}$ in the (3,2,2) scenario}

We found earlier, in Section~\ref{sec:inequivalence}, that the classical
stochastic and deterministic sets $\mathcal{C}$ and
$\mathcal{C}_{\text{det}}$ are always the same if Alice has two equiprobable
inputs. The $(3,2,2)$ setting is therefore the smallest in which we could
hope to find that $\mathcal{C}$ and $\mathcal{C}_{\text{det}}$ are different
even if Alice's inputs are chosen with the same probabilities
($q_{x} = 1/3$). This is indeed what we find for certain values of the
upper bound $G$ that we impose on the guessing probability.

We applied the method we described in the previous subsection to find the
facets of $\mathcal{C}^{+}$ in the $(3,2,2)$ setting with $q_{x} = 1/3$.
In terms of the correlators $E_{xy} = p(1|x,y) - p(2|x,y)$, in addition to
the trivial conditions $\pm E_{xy} \leq 1$ and $G \geq 1/3$ its facets, up to
relabellings of the inputs and outputs, are
\begin{IEEEeqnarray}{rCl}
  \label{eq:C-facet1}
  -E_{11} - E_{12} - E_{21} + E_{22} + E_{31} &\leq& 6 G - 1 ,\\
  \label{eq:C-facet2}
  -E_{11} + E_{31} &\leq& 6 G - 2 .
\end{IEEEeqnarray}
For comparison, the facets of the deterministic version of the set, which we
could call $\mathcal{C}^{+}_{\text{det}}$,
are\footnote{Ref.~\cite{Tavakoli2020} originally inferred these boundaries by
  deriving the facets of $\mathcal{C}_{\text{det}}$ for multiple fixed values
  of $G$ between $1/3$ and $1$. We rederived them here following the analogue
  of the method of the previous subsection applied to deterministic
  communication, treating $G$ as a free variable. This confirms conclusively
  that the boundaries hold for all values of $G$.}
\begin{IEEEeqnarray}{rCl}
  \label{eq:Cdet-facet1}
  -E_{11} - E_{12} - E_{21} + E_{22} + E_{31}  &\leq& 6 G - 1 , \\
  \label{eq:Cdet-facet2}
  -E_{11} - E_{12} - E_{21} + E_{22} + 2 E_{31} &\leq& 12 G - 4 , \\
  \label{eq:Cdet-facet3}
  -E_{11} + E_{31} &\leq& 6 G - 2 .
\end{IEEEeqnarray}
We see here that $\mathcal{C}^{+}$ and $\mathcal{C}^{+}_{\text{det}}$ share two
nontrivial classes of facets. Of these, \eqref{eq:C-facet2} and
\eqref{eq:Cdet-facet3}, which we can rewrite as
\begin{equation}
  \tfrac{1}{3} p(1|31) + \tfrac{1}{3} p(2|11) \leq G ,
\end{equation}
are instances of the constraints \eqref{eq:max-to-linear-ineqs} that we
pointed out apply regardless of the underlying physical theory in
Section~\ref{sec:GboundMessages}. They are not always facets of the sets
$\mathcal{C}$ and $\mathcal{C}_{\text{det}}$ with $G$ fixed due to Alice
having more inputs than Bob has outputs in this setting: in particular they
become redundant if $G$ is larger than $2/3$. The other boundary
\eqref{eq:C-facet1} and \eqref{eq:Cdet-facet1} common to the two sets, by
contrast, is a nontrivial facet of both $\mathcal{C}$ and
$\mathcal{C}_{\text{det}}$ for all $1/3 \leq G \leq 1$.

The only difference between the stochastic and deterministic classical sets
is the class of boundaries \eqref{eq:Cdet-facet2} unique to
$\mathcal{C}_{\text{det}}$. Eq.~\eqref{eq:Cdet-facet2} nontrivially
constrains the correlations for any $G < 5/6$ but becomes redundant for
$G \geq 5/6$. This tells us that $\mathcal{C}$ and $\mathcal{C}_{\text{det}}$
coincide if $G \geq 5/6$ but that $\mathcal{C}_{\text{det}}$ is a strictly
smaller set than $\mathcal{C}$ for $G < 5/6$ in the $(3,2,2)$ setting with
equiprobable priors.

\subsection{Membership testing by linear programming}

While knowing the boundaries of $\mathcal{C}$ is useful for certain purposes,
it is possible to solve the basic problem of testing for membership in
$\mathcal{C}$ without explicitly needing to derive its boundaries. Given a
bound $G$ on the guessing probability, determining whether or not a given
behaviour $p(b|x, y)$ is contained in the corresponding classical set
$\mathcal{C}$ is equivalent to determining whether the pair
$\bigro{p(b|x, y), G}$ is contained in the set $\mathcal{C}^{+}$ that we
introduced in the previous subsection. This amounts to determining whether
$p(b|x,y)$ and $G$ can respectively be expressed as and bounded\footnote{If
  we follow the exact formulation in the previous subsection then, as we
  point out in Appendix~\ref{sec:finite-message-dim}, $\mathcal{C}^{+}$ has
  one conic generator $\bigro{p(b|x, y), G} = (0, 1)$ in addition to its
  vertices which can be added to any point in $\mathcal{C}^{+}$ to increase
  its guessing probability bound component. Eliminating this conic generator
  results in \eqref{eq:G>=avgGvertex} being an inequality.} by averages
\begin{IEEEeqnarray}{rCl}
  \label{eq:pb-xy-avgpvertex}
  p(b|x,y) &=& \sum_{\lambda} p(\lambda) p_{\lambda}(b|x, y) , \\
  \label{eq:G>=avgGvertex}
  G &\geq& \sum_{\lambda}  p(\lambda) G_{\lambda}
\end{IEEEeqnarray}
of the respective components of vertices
$\bigro{p_{\lambda}(b|x, y), G_{\lambda}}$ of $\mathcal{C}^{+}$.

Recalling how we generate the vertices of $\mathcal{C}^{+}$ from those of
$\mathcal{M}^{+}$ in the previous subsection, we may substitute every vertex
probability $p_{\lambda}(b|x, y)$ in \eqref{eq:pb-xy-avgpvertex} by
\begin{equation}
  p_{\lambda}(b|x, y) = \sum_{m} p_{\lambda}(m|x) p_{\lambda}(b|y, m)
\end{equation}
where $p_{\lambda}(m|x)$ is a vertex probability of $\mathcal{M}^{+}$ and
$p_{\lambda}(b|y, m)$ is a deterministic response function. Furthermore, we
may limit the number of messages to an alphabet of size
$n_{M} = 2^{n_{X} - 1}$ without loss of generality. This allows us to express
the problem above as
\begin{IEEEeqnarray}{rCl}
  \label{eq:pbxy-sumlambda}
  p(b|x, y)
  &=& \sum_{\lambda, m} p(\lambda) p_{\lambda}(m|x) p_{\lambda}(b|y, m) , \\
  \label{eq:G-sumlambda}
  G &\geq& \sum_{\lambda}  p(\lambda) G_{\lambda} ,
\end{IEEEeqnarray}
with $p_{\lambda}(b|y, x) \in \{0, 1\}$ and where
$\bigro{p_{\lambda}(m|x), G_{\lambda}}$ is a vertex of $\mathcal{M}^{+}$, for
all $\lambda$.

There are a finite number $n_{K} = n_{B}^{n_{M} \cdot n_{Y}}$ of possible
deterministic response functions on Bob's side. Let us denote these
$p_{k}(b|y, m)$, identified by an index $k$ taking one of
$n_{B}^{n_{M} \cdot n_{Y}}$ distinct values, and group the remaining terms by
$k$. Defining $\Lambda_{k}$ as the set of $\lambda$s appearing in the problem
above for which
\begin{equation}
  p_{\lambda}(b|y, m) = p_{k}(b|y, m) ,
\end{equation}
we can rewrite our problem as
\begin{IEEEeqnarray}{rCl}
  \label{eq:pbxy-sumk}
  p(b|x, y)
  &=& \sum_{k, m} p(k) p_{k}(m|x) p_{k}(b|y, m) , \\
  \label{eq:G-sumk}
  G &\geq& \sum_{k} p(k) G_{k}
\end{IEEEeqnarray}
where
\begin{IEEEeqnarray}{rCl}
  p(k) &=& \sum_{\lambda \in \Lambda_{k}} p(\lambda) , \\
  \label{eq:pkmx-def}
  p_{k}(m|x)
  &=& \sum_{\lambda \in \Lambda_{k}} p(\lambda | k) p_{\lambda}(m|x) , \\
  \label{eq:Gk-def}
  G_{k} &=& \sum_{\lambda \in \Lambda_{k}}  p(\lambda | k) G_{\lambda} ,
\end{IEEEeqnarray}
and $p(\lambda | k)$ is defined in such a way that
$p(k) p(\lambda | k) = p(\lambda)$.

The reexpression \eqref{eq:pbxy-sumk} and \eqref{eq:G-sumk} of our problem is
superficially the same as \eqref{eq:pbxy-sumlambda} and
\eqref{eq:G-sumlambda} except that now there is a known finite number of the
indices $k$ and $m$, while the pairs $\bigro{p_{k}(m|x), G_{k}}$ are no longer
necessarily vertices of $\mathcal{M}^{+}$. The $\bigro{p_{k}(m|x),
  G_{k}}$s are still necessarily \emph{contained} in $\mathcal{M}^{+}$,
however, since $\mathcal{M}^{+}$ is convex, and thus by definition satisfy
\begin{equation}
  \label{eq:pkmx-Gk-constraint}
  G_{k} \geq \sum_{m} \max_{x} q_{x} p_{k}(m|x)
\end{equation}
together with
\begin{IEEEeqnarray}{c+c}
  \label{eq:pkmx-pos-and-norm}
  p_{k}(m|x) \geq 0 , & \sum_{k} p_{k}(m|x) = 1 .
\end{IEEEeqnarray}
Using these constraints in place of \eqref{eq:pkmx-def} and \eqref{eq:Gk-def}
and then eliminating the $G_{k}$s simplifies the problem to
\begin{IEEEeqnarray}{rCl}
  p(b|x, y) &=& \sum_{k, m} p(k) p_{k}(m|x) p_{k}(b|y, m) , \\
  G &\geq& \sum_{k, m} p(k) \max_{x} q_{x} p_{k}(m|x) ,
\end{IEEEeqnarray}
where $p(k)$ and $p_{k}(m|x)$ are probability distributions.

To turn this into a linear programming problem we combine $p(k)$ and
$p_{k}(m|x)$ into a joint distribution,
\begin{equation}
  p(k, m|x) = p(k) p_{k}(m|x) ,
\end{equation}
which satisfies the marginal condition that $\sum_{m} p(k, m|x) = p_{k}$ is
independent of $x$ for all $k$. With this last replacement the full problem
becomes
\begin{IEEEeqnarray}{rCl}
  \label{eq:lp-pbxy}
  p(b|x, y) &=& \sum_{k, m} p(k, m|x) p_{k}(b|y, m) , \\
  \label{eq:lp-G}
  G &\geq& \sum_{k, m} \max_{x} q_{x} p(k, m|x) , \\
  \label{eq:lp-positivity}
  p(k, m|x) &\geq&0 , \\
  \label{eq:lp-normalisation}
  \sum_{k, m} p(k, m|x) &=& 1 , \\
  \label{eq:lp-marginal}
  \sum_{m} p(k, m|x) &=& \sum_{m} p(k, m|x') , \qquad \forall x \neq x' .
\end{IEEEeqnarray}
Recalling that \eqref{eq:lp-G} is a shorthand for $n_{M}^{n_{B}}$ linear
inequalities, determining whether there exist $n_{K} \cdot n_{M} \cdot n_{X}$
weights $p(k, m|x)$ that satisfy
Eqs.~\eqref{eq:lp-pbxy}--\eqref{eq:lp-marginal} for a given behaviour
$p(b|x, y)$ is a linear programming feasibility problem.

We remark, finally, that if we drop the marginal constraint
\eqref{eq:lp-marginal} and combine $(k, m)$ into a new variable which we
rename $m$, we recover the definition of the classical set $\mathcal{C}$ that
we started with in Section~\ref{sec:GboundMessages}. This confirms that we
did not inadvertently relax the problem when we replaced the vertices
$\bigro{p_{\lambda}(m|x), G_{\lambda}}$ of $\mathcal{M}^{+}$ with the
conditions \eqref{eq:pkmx-Gk-constraint} and \eqref{eq:pkmx-pos-and-norm} on
$p_{k}(m|x)$. Deriving the linear programming feasibility problem following
our characterisation of $\mathcal{C}^{+}$, however, allows us to put a finite
upper limit $n_{K} \cdot n_{M} \cdot n_{X}$, with
$n_{K} = n_{B}^{n_{M} \cdot n_{Y}}$ and $n_{M} = 2^{n_{X} - 1}$, on the
number of weights $p(k, m|x)$ that we need to consider.

\section{Characterising quantum correlations}
\label{sec:tools}

In this section, we develop tools for the characterisation of informationally restricted quantum correlations. In Section~\ref{sectionseesaw}, we develop an efficient method for optimising any given linear witness from inside the set of informationally restricted quantum correlations $\mathcal{Q}$. Hence, this method enables lower bounds on quantum correlations. In Section~\ref{subsec:upperbnds}, we present a hierarchy of semidefinite relaxations of $\mathcal{Q}$ (and of $\mathcal{Q}_{\text{pure}}$). This allows us to establish increasingly precise necessary criteria of a given correlation admitting a quantum model. In Section~\ref{sectionQexample}, we apply these methods to the simplest relevant communication experiment and use it to device-independently quantify the information content of a quantum ensemble. In Section~\ref{sectionQresource}, we focus on the case of one bit of information and prove several strict resource inequalities involving two-dimensional systems, pure-state informationally restricted systems and general informationally restricted systems, in both the quantum and classical setting.

\subsection{Lower bounds: alternating convex search method}
\label{sectionseesaw}

In many situations arising in the study of quantum correlations, it is possible to use alternating convex searches in order to optimise a linear functional of the quantum correlations (a linear ``witness''), such as in the case of Bell inequalities \cite{Werner2001,Pal2010} or quantum dimension witnesses \cite{Magic7}. Such a search amounts to attempting to solve the full optimisation problem (over both states and measurements) by repeatedly optimising over the states and measurements separately in an alternating manner. The advantage of such an approach is that often each separate optimisation, over states (measurements) for fixed measurements (states), is convex and can be solved by standard methods. While alternating convex search often works well in practice, it is not guaranteed to converge and therefore only offers lower bounds on the optimal quantum correlations.

In order to optimise a linear witness over the set of informationally restricted quantum correlations, one encounters a less straightforward situation. For a fixed set of states, it is clear that the optimisation over the set of measurements can be evaluated as a semidefinite program (SDP). In contrast, for a fixed set of measurements, the optimisation over the set of states is less obvious due to the relevance of the informational restriction. Evidently, the optimisation must be performed under the constraint $P_{\guess} \leq G $ which itself involves a maximisation over the extraction POVM $\{N_x\}_x$. We show how this difficulty can be overcome so that lower bounds on informationally restricted quantum correlations can be efficiently computed through alternating implementations of SDPs.

Consider that we are given a linear witness $\mathcal{A}$, in general written as 
\begin{equation}
  \mathcal{A}=\sum_{x,y,b}c_{xyb}p(b|x,y)=\sum_{x,y,b}c_{xyb}\Tr \bigsq{\rho_{x} M_{b|y}}
\end{equation}
for some real coefficients $c_{xyb}$, and asked to maximise it over the set of informationally restricted states $\rho_x$ and measurements $M_{b|y}$. For this purpose, let us define an auxiliary positive semidefinite operator $\sigma$ with the property that 
\begin{equation}\label{sigma}
	\forall x: \quad \sigma \geq q_x \rho_x.
\end{equation}
This allows us to place the following upper bound on the guessing probability:
\begin{IEEEeqnarray}{rCl}
  \label{guessbound}
  P_{\guess}(X|B) &=& \max_{\{N_x\}_x} \sum_x q_x\Tr \bigsq{\rho_x N_x} \nonumber \\
  &\leq& \max_{\{N_x\}_x} \sum_x \Tr \bigsq{\sigma N_x} = \Tr[\sigma],
\end{IEEEeqnarray}
where we have used that $\sum_x N_x=\openone$. The introduction of $\sigma$ stems from considering the semidefinite dual of the guessing probability and does therefore not constitute a relaxation of the problem \cite{Watrous}. Its advantage is that it allows us to treat the informational restriction as a tracial constraint enforced through the additional semidefinite constraints in \eqref{sigma}. We may therefore cast the maximisation of the linear witness $\mathcal{A}$, for a given bound $G$ on the guessing probability, as the following optimisation problem:
\begin{IEEEeqnarray}{l+rCl+rCl}
  \max_{\rho_x,\sigma,M_{b|y}} & \IEEEeqnarraymulticol{6}{l}{
    \sum_{x,y,b}c_{xyb}\Tr \bigsq{\rho_x M_{b|y}}}
  \nonumber \\
  \text{such that} & \rho_x &\geq& 0 , & \Tr[\rho_x] &=& 1, \nonumber \\
  & \sigma &\geq& q_x \rho_x, & \Tr[\sigma] &\leq& G, \nonumber\\
  & M_{b|y} &\geq& 0, & \sum_b M_{b|y} &=& \openone . \label{eq:poloptim}
  \IEEEeqnarraynumspace
\end{IEEEeqnarray}
If we fix the measurement operators $\{M_{b|y}\}$, this problem becomes an SDP for the states $\{\rho_x\}$ and $\sigma$. Conversely, if we fix the states $\{\rho_x\}$ and $\sigma$, it is an SDP for the measurement operators $\{M_{b|y}\}$. We can thus alternate these SDPs to obtain a lower bound on the optimal value. Note that it is implicit that these SDPs must be performed in a given Hilbert space dimension, but one may find successively better lower bounds by increasing the dimension. The usefulness of this method is exemplified in Section~\ref{sectionQexample}.

Note that the above approach cannot be applied to the pure-state set $\mathcal{Q}_{\text{pure}}$, since the condition $\rho_x\geq 0$ would have to be replaced by $\rho_x^2=\rho_x$, which is nonlinear. The existence of a practical algorithm lower-bounding the general quantum set $\mathcal{Q}$ is another advantage of our general formulation.

\subsection{Upper bounds: hierarchy of semidefinite relaxations}\label{subsec:upperbnds}
The idea used in the previous section, of introducing the auxiliary operator $\sigma$, can be further leveraged to systematically obtain increasingly precise upper bounds on the informationally restricted set of quantum correlations. We now present a hierarchy of semidefinite relaxations for the set $\mathcal{Q}$, which is based on the tracial variant \cite{burgdorf_tracial_2013,klep_constrained_2016} of the NPA non-commutative polynomial optimisation hierarchy \cite{Navascues2008,Pironio2010b}. 

Let us first slightly rewrite the problem \eqref{eq:poloptim} as
\begin{IEEEeqnarray}{l+l}
    \max_{\rho_x,\sigma,M_{b|y}} & 
      \sum_{x,y,b}c_{xyb}\Tr \bigsq{\rho_x M_{b|y}}
    \nonumber \\
    \text{such that} & \rho_x -\rho_x^2\geq 0,\quad  \Tr[\rho_x] = 1, \nonumber \\  \label{eq:poloptim2}
    & \sigma - q_x \rho_x\geq 0, \quad G\openone-\sigma\geq 0, \quad\Tr[\sigma] \leq G, \nonumber\\
    &  M_{b|y} M_{b'|y} = \delta_{bb'} M_{b|y},\quad \sum_b M_{b|y} = \openone ,  \IEEEeqnarraynumspace
  \end{IEEEeqnarray}
where compared with \eqref{eq:poloptim}, we have replaced the constraint $\rho_x\geq 0$ by $\rho_x-\rho_x^2\geq 0$, added the redundant constraint $G\openone-\sigma\geq 0$, and assumed, without loss of generality if we do not bound the dimension $d$ of the Hilbert space, that the measurements $\{M_{b|y}\}_b$ are projective. The optimization problems \eqref{eq:poloptim} and \eqref{eq:poloptim2} are entirely equivalent, but the second formulation is better suited for the tracial non-commutative optimization method of \cite{burgdorf_tracial_2013}\footnote{Specifically, the constraints $\rho_x-\rho_x^2$ imply not only that $\rho_x\geq 0$ but also that $\rho_x\leq \openone$. Together with  the constraint $\sigma\leq G\openone$ this guarantees that the feasible set of \eqref{eq:poloptim2} satisfies the archimedean assumption and that the entries of the moment matrices stay bounded. Assuming that the measurements are projectives instead of general POVMs dispenses us from introducing localizing matrices associated with them.}, which we now explain how to apply.

Let $w$ denote a monomial, i.e. a product, of the $n_X+1+n_Bn_Y$ basic operators $\rho_1,\rho_2,\dotsc,$ $\rho_{n_X}$, $\sigma$, and
$M_{1,1},\dotsc, M_{n_B|1},M_{1|2}$, $\dotsc, M_{n_B|n_Y}$. We refer to the number $k$ of such basic operators in the product $w$ as the degree
$k$ of this monomial. By convention, the identity operator $\openone$ is the
monomial of degree $0$. Let $\mathcal{W}_k$ denote the set of all monomials
of degree at most $k$ and let $n(k)$ denote the number of such
monomials. Linear combinations $p=\sum_{w\in\mathcal{W}_k} \alpha_w w$ of the
monomials then correspond to polynomials of degree $k$ in the basic operators.

Let $L$ be a linear functional that assigns to each  monomial $w$ in $\mathcal{W}_{2k}$ of degree $2k$ the real number $L(w)$, and thus which assigns to each polynomial $p=\sum_{W\in\mathcal{W}_{2k}} \alpha_w w$ of degree $2k$ the real number $L(p)=\sum_{w\in\mathcal{W}_k} \alpha_w L(w)$. Given such a functional $L$, we define
\begin{itemize}
  \item the \emph{moment} matrix $\Gamma_k(L)$, as the matrix of size $n(k)$ whose entries are indexed by monomials $u,v\in \mathcal{W}_k$ and are equal to $[\Gamma_k(L)]_{u,v}=L(u^\dagger v)$;
  \item the \emph{localizing} matrix $\Gamma_k(L;p)$ associated to a polynomial $p$ of degree two or less, as the matrix of size $n(k)-1$ whose entries are indexed by monomials $u,v\in \mathcal{W}_{k-1}$ and are equal to $[\Gamma_k(L;p)]_{u,v}=L(u^\dagger p\,v)$.
\end{itemize}

Consider now the following problem for $k\geq 1$,
\begin{IEEEeqnarray}{ll}
  \max_{L} & 
    \sum_{x,y,b}c_{xyb}\,L\left(\rho_x M_{b|y}\right)
      \nonumber \\
  \text{such that } & \Gamma_k(L)\geq 0,\nonumber\\
  & \Gamma_k(L;\rho_x -\rho_x^2)\geq 0,\quad  L(\rho_x) = 1, \nonumber \\  \label{eq:gamma-eq}
  & \Gamma_k(L;\sigma - q_x \rho_x)\geq 0, \quad \Gamma_k(L;G\openone-\sigma)\geq 0, \nonumber\\
  &\quad   L(\sigma) \leq G, \nonumber\\
  &  L(p)=L(p'), \text{ if $\Tr[p] = \Tr[p']$} \\
  & \qquad \text{for any polynomials  $p, p'$ of degree $2k$,} \nonumber
  %\IEEEeqnarraynumspace
\end{IEEEeqnarray}
where in the last condition the identity $\Tr[p]=\Tr[p']$ is evaluated by taking into account the polynomial identities $M_{b|y} M_{b'|y} = \delta_{bb'}M_{b|y}$ and $ \sum_b M_{b|y} = \openone$ satisfied by the measurement operators. This optimization problem is an SDP (since it amounts to optimize $n(2k)$ variables, the values $L(w)$ of the monomials $w$ of degree less than $2k$, subject to linear constraints and to the positivity of matrices whose entries are linearly related to these variables).

Clearly any solution of \eqref{eq:poloptim2} defines a solution of \eqref{eq:gamma-eq} through $L(w) = \Tr[w]$ \footnote{This can be seen by following the same lines as in \cite{Navascues2008,Pironio2010b}, where the linear functional was instead defined as $L(w)=\bra{\Psi} w \ket{\Psi}$ for some reference state $\ket{\Psi}$.}. Thus the problem \eqref{eq:gamma-eq} represents an SDP relaxation of \eqref{eq:poloptim2} approximating the set $\mathcal{Q}$ from the outside.  
By increasing the relaxation level $k$, one obtains a hierarchy of increasingly constraining conditions on $\mathcal{Q}$. 

Note that the above method can also be used to characterise the set of pure-state quantum correlations $\mathcal{Q}_{\text{pure}}$ by replacing in \eqref{eq:poloptim} the positivity constraints $\rho_x-\rho_x^2 \geq 0$ by the polynomial constraints $\rho_x=\rho_x^2$, resulting in the simpler relaxation
\begin{IEEEeqnarray}{ll}\label{eq:gamma-eq-pure}
  \max_{L} & 
    \sum_{x,y,b}c_{xyb}\,L\left(\rho_x M_{b|y}\right)
      \nonumber \\
  \text{such that } & \Gamma_k(L)\geq 0,\nonumber\\
  & \Gamma_L(\rho_x) = 1, \nonumber \\  \label{eq:gamma-eq2}
  & \Gamma_k(L;\sigma - q_x \rho_x)\geq 0, \quad \Gamma_k(L;G\openone-\sigma)\geq 0, \nonumber\\
  &\quad   L(\sigma) \leq G, \nonumber\\
  &  L(p)=L(p'), \text{ if $\Tr[p] = \Tr[p']$} \\
  & \qquad \text{for any polynomials  $p, p'$ of degree $2k$,} \nonumber
  %\IEEEeqnarraynumspace
\end{IEEEeqnarray}
where the last condition is evaluated using, in addition to the polynomial constraints on the measurement operators, the conditions $\rho_x=\rho_x^2$. 

We remark that by additionally imposing that all operators commute, we can also bound classical correlations via the above SDPs. This can be useful in scenarios that are too large to be efficiently treated with the methods developed in Section~\ref{sec:classical}.

Finally, let us stress that the series of SDP relaxations that we introduced are relaxations. Convergence to the exact quantum set is not guaranteed in the limit $k\rightarrow\infty$, see \cite{burgdorf_tracial_2013,klep_constrained_2016,gribling_lower_2019} for more details about the general properties of the SDP hieararchy for non-commutative tracial optimization.

\subsection{Device-independent witnessing of the information content of quantum communication}
\label{sectionQexample}

Consider a quantum communication experiment in which we do not know the amount of information communicated from Alice to Bob. Is it possible to determine a lower bound on the amount of information that Alice must send to Bob given only the observed correlations $p(b|x,y)$? This amounts to the task of device-independently 	testing the information content of quantum communication. Using the tools of the previous sections, we exemplify such device-independent certification in the simplest relevant communication experiment.

As we have seen in Section~\ref{sec:stochastic-same}, there can be no quantum
advantage when the scenario only features two states. Moreover, no advantage is possible when Bob only has a single input, because his measurement could then be performed already in Alice's lab and the outcomes simply relayed to Bob as classical communication (since performing a measurement cannot increase the guessing probability). Therefore, the simplest
relevant scenario in which we expect a quantum advantages is that in which
Alice has three states $(n_X=3)$ and Bob has two binary-outcome measurements
$(n_Y=n_B=2)$. In this scenario, we focus on the linear witness
\eqref{eq:C-facet1} (here labelled $\mathcal{A}_{322}$) corresponding to a
facet of the classical polytope under uniform priors ($q_x=1/3$).

\begin{figure}
  \centering
  \includegraphics{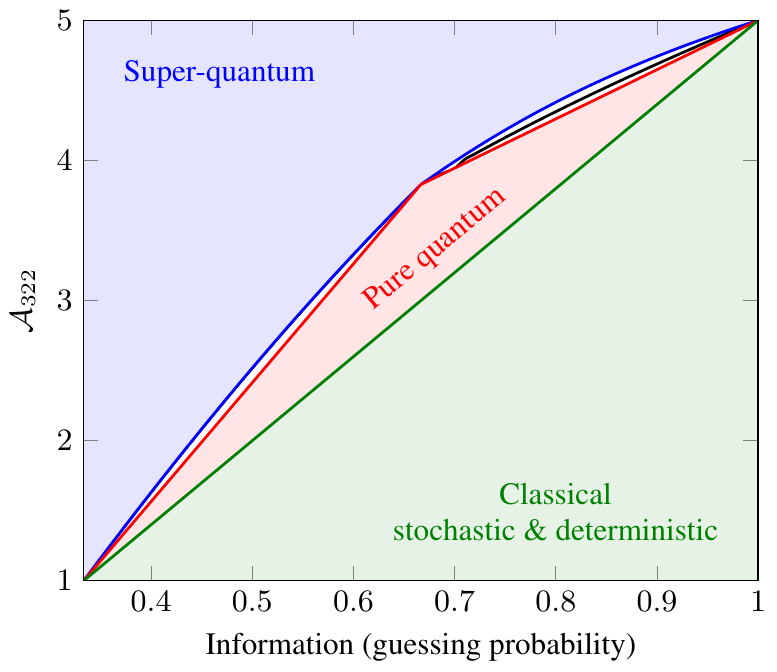}
  \caption{The witness $\mathcal{A}_{322}$ versus the information (in terms of the guessing probability). The plot displays an upper bound (blue) and lower bound (black) on general quantum models, a tight upper bound on pure-state quantum models (red) and a tight upper bound on classical models (green). As the first two curves coincide in the interval $P_{\guess}\in[\frac{1}{3},\frac{2}{3}]$, this part of the quantum boundary is fully characterised. However, in the interval $P_{\guess}\in(\frac{2}{3},1]$, the quantum boundary is not fully characterised but delimited by the blue and black curves.\label{Fig322}}
\end{figure}

Firstly, we apply the SDP hierarchy for the set $\mathcal{Q}_{\text{pure}}$ to find upper bounds on $\mathcal{A}_{322}$ as a function of the guessing probability (information). We implemented the SDP relaxation \eqref{eq:gamma-eq-pure} with $k=3$ but to simplify the numerical optimisation considered a subset of all SDP and linear constraints. Specifically, we only imposed the positivity of the submatrix of $\Gamma_3(L)$ whose rows and columns are indexed by the monomials
\begin{multline}
  \{ \openone,\, \sigma,\, \rho,\, M,\, \rho M,\,  \rho\rho,\, MM,\, \rho \sigma,\, M\sigma,\, \rho\rho\rho, \\
	MM\sigma,\, \rho M M,\, \rho M \sigma,\, \rho M \rho \} ,
\end{multline}
the positivity of the submatrices $\Sigma^x_2(L)$ whose rows and columns are  indexed by the monomials
\begin{equation}\label{eq:mon-list}
\{\openone,\, \rho,\, M,\, \rho\rho,\, MM,\, \rho M\} ,
\end{equation}
and the linear constraints $L(P)=L(P')$ involving the entries of such matrices. 
This corresponds to a $98\times98$ moment submatrix $\Gamma$ and three $25\times 25$ localising submatrices $\Sigma^x$. Evaluating the corresponding SDPs for different informational restrictions, we obtain the red curve illustrated in Fig.~\ref{Fig322}. Notably, this upper bound is in fact tight, since it coincides with the explicit pure-state quantum strategy reported in Ref.~\cite{Tavakoli2020} (thus proving its optimality).

Similarly, we have also implemented the SDP hierarchy for the general quantum set $\mathcal{Q}$ using submatrices of the localising matrices $P^x_3(L)$ based on the same monomial list $\eqref{eq:mon-list}$ as for $\Sigma^x_3(L)$. The obtained bounds on the witness are given by the blue curve in Fig.~\ref{Fig322}. We observe that for every guessing probability $P_{\guess}\in(\frac{1}{3},1)\setminus \{\frac{2}{3}\}$ we find a larger bound in the general setting as compared to the pure-state setting. In order to show that this gap is not an artefact of the bounds in the general setting not being tight, we have employed the alternating convex search described in Section~\ref{sectionseesaw} to construct explicit quantum models. The obtained values of the witness are illustrated by the black curve in Fig.~\ref{Fig322}. We find that for $P_{\guess}\in[\frac{1}{3},\frac{2}{3}]$, the upper and lower bounds in the general setting accurately coincide. In the interval $P_{\guess}\in(\frac{2}{3},1)$ a small gap between the upper and lower bound remains. Nevertheless, our lower bounds  exceed the upper bounds for the pure-state setting, thus proving that informationally restricted quantum correlations outperform their pure-state counterparts. It is interesting to note that for the special case of $P_{\guess}=\frac{2}{3}$, which corresponds precisely to $\mathcal{I}(X|B)=1$ bit of information, there is no discrepancy between $\mathcal{Q}$ and $\mathcal{Q}_{\text{pure}}$.

We can interpret these results in the context of device-independent tests of information. If the information content of the quantum communication is not known, then we may use the upper bound on the quantum correlations (blue curve) to determine a bound on the minimal amount of information required to explain the observed correlations in a quantum model. For example, Ref.~\cite{Ahrens2012} experimentally implemented this communication experiment using both qubit and qutrit ensembles and reported a witness value of $\mathcal{A}^{\text{qubit}}_{322}=3.7815\pm 0.0782$ and $\mathcal{A}^{\text{qutrit}}_{322}=4.9303\pm 0.1032$ respectively. In order to determine the information content of these ensembles (without assuming their respective dimensions), we use our upper bounds on the quantum correlations. Specifically, when the experimental errors are taken into consideration, we certify a quantum information content of at least $\mathcal{I}(X|B)=0.98 \pm 0.02 $ bits for the first ensemble and $\mathcal{I}(X|B)=1.54\pm 0.05$ bits for the second ensemble. Both these results nearly saturate the  maximal possible information content of qubit and qutrit ensembles, namely $1$ bit and $\log_{2} 3$ bits respectively.

\subsection{Resource inequalities for one bit of information}\label{sectionQresource}

Consider the information restriction $\mathcal{I}(X|B)\leq \alpha$ with $\alpha=\log_2 d$ for some integer $d\geq 2$. This is a particularly interesting case since it enables a meaningful comparison of classical and quantum correlations to those that can be obtained from $d$-dimensional classical and quantum communication. Here, we focus on the simplest case of $d=2$ ($\mathcal{I}_X\leq 1$ bit) and consider the comparative relation between classical and quantum correlations respectively when obtained from i) communication of two-level systems, ii) one bit of communication in pure-state models and iii) one bit of communication in general models. Let us denote the set of classical and quantum correlations achievable with two-dimensional communication by $\mathcal{C}_{\text{dim}}$ and $\mathcal{Q}_{\text{dim}}$. It is clear that the following two chains of inclusions must be true:
\begin{IEEEeqnarray}{c+t+c}
  \label{setchain}
  \mathcal{C}_{\text{dim}} \subseteq \mathcal{C}_{\text{det}} \subseteq \mathcal{C}
  &and&
  \mathcal{Q}_{\text{dim}} \subseteq \mathcal{Q}_{\text{pure}} \subseteq \mathcal{Q}. \IEEEeqnarraynumspace
\end{IEEEeqnarray}
The first inclusion in each case follows from the fact that every ensemble of classical or quantum two-level systems can be simulated by classical or quantum ensembles of pure two-level systems under shared randomness\footnote{Recall that since every ensemble of two-level systems carries no more than one bit of information, then also their mixture under shared randomness does not lead to more than one bit of information.}.	 The second inclusion on each line follows trivially from the fact that general classical and quantum models admit deterministic and pure-state models respectively as special cases.

It is interesting to determine which of the inclusions \eqref{setchain} are strict, i.e., which classical and quantum resources are fundamentally different. We first focus on the quantum case and prove that all three resources are inequivalent. Notably, Ref.~\cite{Tavakoli2020} proved that $\mathcal{Q}_{\text{dim}} \subset \mathcal{Q}$ using a construction that involved 16 states. The proofs presented here are simpler, as they only require three states, but inherently different as they are based on biasing the prior probabilities.

Consider again the input/output scenario $(n_X,n_Y,n_B)=(3,2,2)$ and once again the witness $\mathcal{A}_{322}$. In the previous section, we saw that for $\mathcal{I}_X\leq 1$ bit ($P_{\guess}\geq \frac{2}{3}$), there was no discrepancy between the general quantum model and the pure-state quantum model. In addition, if we restrict to qubits, the witness $\mathcal{A}_{322}$ reduces to that introduced in Ref.~\cite{Gallego2010}, whose maximum is known again to give the same result. However, consider now that we change the prior distribution of Alice's inputs: instead of being uniform, let us choose it as $q_1=q_2=\frac{2}{5}$ and $q_3=\frac{1}{5}$. Since $H_\mathrm{min}(X)=\log_{2}(5)-1$, the guessing probability corresponding to one bit of information is $P_{\guess}=\frac{4}{5}$. What now are the largest possible values of $\mathcal{A}_{322}$ under qubits, pure-state models with $P_{\guess}\leq \frac{4}{5}$, and general models with $P_{\guess}\leq \frac{4}{5}$? 

Since biasing the prior affects the information constraint but not the dimension of the physical system, it follows that the largest value of $\mathcal{A}_{322}$ remains unaffected when evaluated over qubits. We have
\begin{equation}\label{qq}
  \mathcal{A}_{322}\stackrel{\mathcal{Q}_{\text{dim}}}{\leq} 1+2\sqrt{2}\approx 3.8284,
\end{equation}
which is a tight bound. However, in the case of pure-state models and general quantum models, biasing the prior means that Bob already has some knowledge of Alice's input. Thus, we would intuitively expect that the correlations improve as compared to the unbiased case. This intuition can be proven using the tools from the previous sections. Evaluating the respective semidefinite relaxations of the set of quantum correlations, we find that 
\begin{IEEEeqnarray}{c+c}
  \mathcal{A}_{322} \stackrel{\mathcal{Q}_{\text{pure}}}{\leq} 4.3184 , &
  \mathcal{A}_{322} \stackrel{\mathcal{Q}}{\leq} 4.4641 .
\end{IEEEeqnarray}
We use alternating convex search to place a lower bound on the witness in the stochastic case: for qubits we achieve $\mathcal{A}_{322}=3.8284$ (saturating \eqref{qq}), for qutrits we achieve $\mathcal{A}_{322}=4.2641$ and for ququarts we achieve $\mathcal{A}_{322}=4.4142$. The ququart strategy uses one pure state and two mixed states each with spectra $(1/2,1/2,0,0)$. The lower bound obtained with ququarts is sufficient to outperform pure-state quantum models and conclude that $\mathcal{Q}_{\text{pure}}\subset \mathcal{Q}$. Moreover, in order to also show that  $\mathcal{Q}_{\text{dim}}\subset \mathcal{Q}_{\text{pure}}$, it is sufficient to note that the following strategy based on pure-state quantum communication outperforms the qubit bound. Let Alice prepare the qutrit states 
\begin{IEEEeqnarray}{c+c+c}
  \ket{\psi_1} = \frac{1}{2} \begin{pmatrix}
    \sqrt{3} \\ 1 \\ 0
  \end{pmatrix} , &
  \ket{\psi_2} = \frac{1}{2} \begin{pmatrix}
    1 \\ \sqrt{3} \\ 0
  \end{pmatrix} , &
  \ket{\psi_3} = \begin{pmatrix}
    0 \\ 0 \\ 1
  \end{pmatrix} . \IEEEeqnarraynumspace
\end{IEEEeqnarray}
It is easily checked (e.g.~via an SDP) that the guessing probability is $P_{\guess}=4/5$. Then, let Bob perform compatible measurements $\{\ket{3},\ket{1+2}\}$ and $\{\ket{2},\ket{1+3}\}$. Then, one finds $\mathcal{A}_{322}=4$ which exceeds the qubit bound.

Let us now consider the same problem with classical resources. Using the tools from Section~\ref{sec:classical}, we can straightforwardly show the tight inequalities
\begin{IEEEeqnarray}{c+c+c}
  \mathcal{A} \stackrel{\mathcal{C}_{\text{dim}}}{\leq} 3 , &
  \mathcal{A} \stackrel{\mathcal{C}_{\text{det}}}{\leq} 4 , &
  \mathcal{A} \stackrel{\mathcal{C}}{\leq} 4 ,
\end{IEEEeqnarray}
which immediately assert that informationally restricted classical
correlations are more powerful than dimensionally restricted classical
correlations; specifically $\mathcal{C}_{\text{dim}}\subset \mathcal{C}_{\text{det}}$. However, it still does not determine whether $\mathcal{C}_{\text{det}}$ is a strict subset of $\mathcal{C}$ for one bit of information. This is left as an open problem.

\section{Semi-device-independent random number generation}

In the previous section, we have seen how quantum correlations can be bounded in communication experiments in which the only assumption is a bound on the amount of information that the communication carries. Here, we leverage these methods towards application in semi-device-independent RNG. In a first example, we focus on the facet-defining witness $\mathcal{A}_{322}$ and compute the certified randomness as a function of the information. This allows us to obtain a nearly optimal RNG rate. In a second example, we consider the case of one bit of information and consider the amount of randomness that can be robustly certified under a conventional qubit assumption as compared to that certified under an information assumption. We show that the correlations used in a standard qubit experiment can be recycled to certify the same amount of randomness when the assumption is relaxed to the strictly weaker information assumption.

\subsection{Randomness versus information}

\begin{figure}[t!]
  \centering
  \includegraphics{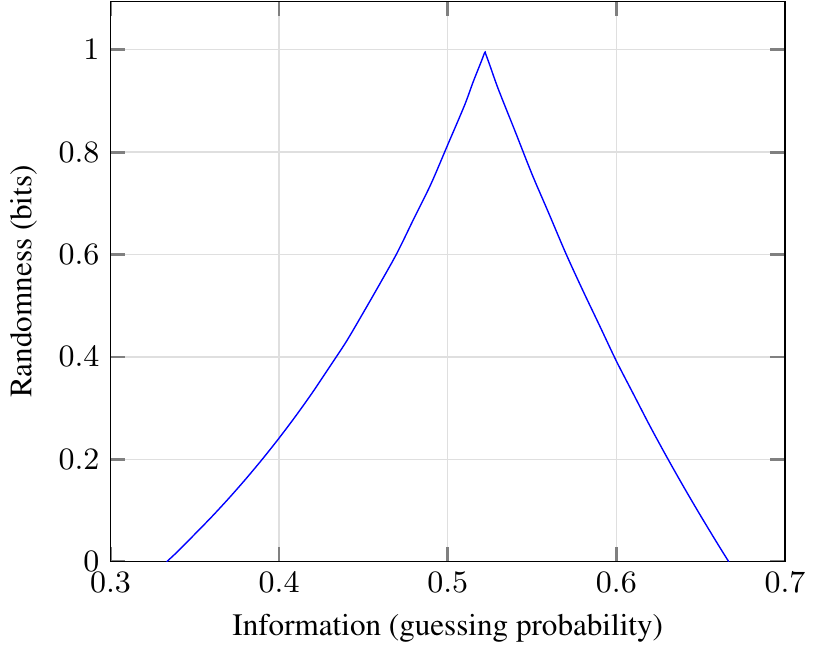}
  \caption{Randomness versus the information (quantified via the guessing probability) of Alice's communication. The results are obtained for the maximal quantum value of the witness $\mathcal{A}_{322}$ in general quantum communication models.\label{RNGvsInfo}}
\end{figure}

Let us again consider the witness $\mathcal{A}_{322}$ in (a general) quantum model. In Section~\ref{sectionQexample} we obtained the maximal quantum witness value for any information between zero and one bit, corresponding to a guessing probability $P_{\guess}\in[\frac{1}{3},\frac{2}{3}]$. Here, we evaluate the extractable randomness in the output of Bob associated to such maximal quantum witness values. Specifically, we consider that Alice and Bob decide to extract randomness from the event corresponding to Alice's third input ($x=3$) and Bob's first input ($y=1$). Then, the certified randomness is given by the min-entropy $H_\text{min}=-\log_2 p_*$, where $p_*=\max\{p(1|3,1),p(2|3,1)\}$, compatible with the observed maximal value of $\mathcal{A}_{322}$\footnote{To enhance numerical feasibility, we only impose the optimal value of $\mathcal{A}_{322}$ up to four decimals.}. Using the introduced semidefinite relaxations, we can place an upper bound on $p_*$ which translates into a lower bound on the certified randomness. The results are illustrated in Fig.~\ref{RNGvsInfo}. These results can also be accurately matched with upper bounds on the randomness obtained via the alternating convex search method (see Section~\ref{sectionseesaw}). Hence, the bound on the certified randomness is tight (up to solver precision). In Fig.~\ref{RNGvsInfo}, we see that by suitably tuning the information in Alice's communication, one can obtain nearly one bit of randomness (which is algebraically maximal for binary-outcome measurements). Specifically, at $P_{\guess} \approx 0.522$ we certify approximately $0.995$ bits of randomness. Hence, we conclude that nearly optimal randomness can be certified under the information assumption. Notably, for $P_{\guess} \approx \frac{2}{3}$, the randomness vanishes. This is due to our choice of setting ($x=3$, $y=1$). A substantial amount of randomness can be certified for $P_{\guess}\approx \frac{2}{3}$ by instead considering the event $(x,y)=(1,1)$. However, the rate is significantly lower than that obtained at the optimal choice of information for $(x,y)=(3,1)$.

\subsection{Qubits versus one bit of information}

We investigate the comparison between certified randomness under the conventional assumption of qubits and our assumption of informational restriction. This comparison is only meaningful for one bit of information; to which we  therefore restrict ourselves. To this end, we focus on a witness that has previously been employed for RNG in dimension bounded systems \cite{HongWei, HongWei2}, namely a quantum random access code.

In a quantum random access code, Alice receives one of four possible inputs labelled by two bits $x = x_1 x_2 \in \{1,2\}^{\times 2}$ while Bob has two possible inputs $y \in \{1, 2\}$ and two possible outputs $b \in \{1, 2\}$. The correlation witness is defined as
\begin{equation}
  \mathcal{A}_{\text{RAC}} = \frac{1}{8} \sum_{x,y} (-1)^{x_y} E_{xy} .
\end{equation}
We analyse this witness in two scenarios, i) Alice sends qubits to Bob (dimension assumption) and ii) Alice sends at most one bit of information to Bob (information assumption). Naturally, since all qubit ensembles carry at most one bit of information, while many higher dimensional ensembles also carry no more than one bit of information, the information assumption is less restrictive than the dimension assumption. It is well known that the optimal value of $\mathcal{A}_{\text{RAC}}$ using qubits is $\frac{1}{\sqrt{2}}$ \cite{Kaniewski2018}. Using the tools from Section~\ref{sec:tools}, we find that $\mathcal{A}_{\text{RAC}}=\frac{1}{\sqrt{2}}$ also is the largest possible value under one bit of information.

\begin{figure}[t!]
  \centering
  \includegraphics{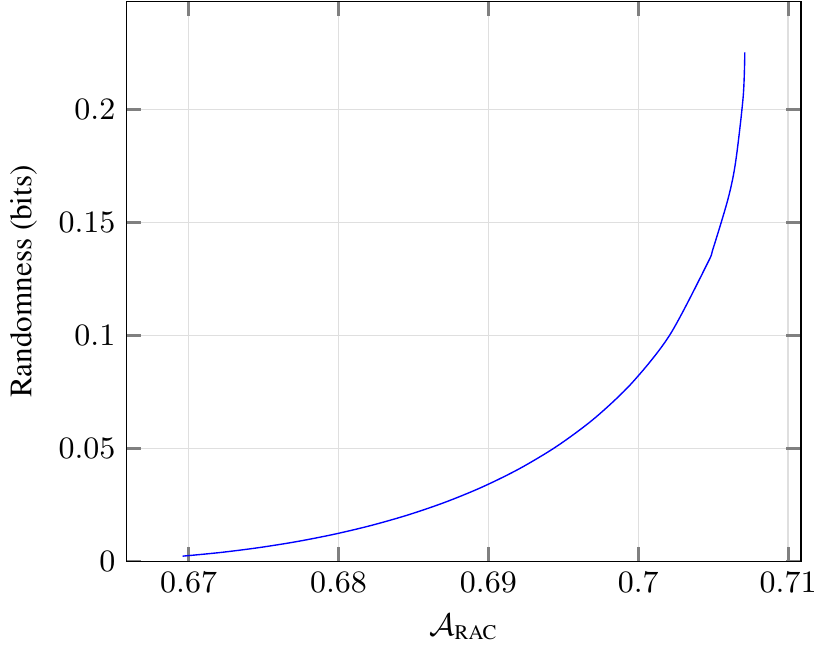}
  \caption{The randomness certified in a quantum random access code. Up to
    numerical precision, the amount of randomness certified under the qubit
    assumption and one bit of information assumption is
    identical.\label{FigRNGQRAC}}
\end{figure}

Due to the symmetries of the witness $\mathcal{A}_{\text{RAC}}$, the choice of event from which randomness is extracted does not influence the amount of randomness certified. We therefore choose the event $(x,y)=(1,1)$ and employ semidefinite relaxations for informationally restricted quantum correlations to place a lower bound on the randomness as a function of the witness. The results are illustrated in Fig.~\ref{FigRNGQRAC}. A nearly optimal value of the witness certifies over $0.2$ bits of randomness while also significantly sub-optimal witness values permit a non-zero amount of certified randomness. Then, we consider the same problem under the assumption of qubit communication. To this end, we have used the symmetrised semidefinite relaxation hierarchy of Refs.~\cite{Navascues2015, Renou2019}. Up to solver precision, we certify the same amount of randomness as is obtained under the information assumption, i.e.~the curve is identical to that displayed in Fig.~\ref{FigRNGQRAC}. Moreover, the obtained lower bounds on the randomness are optimal since we can saturate them with an explicit family of quantum models based on qubits. Hence, we conclude that the quantum random access code allows us to certify the same amount of randomness under the strictly weaker assumption of informational restriction as compared to the dimension bounded scenario, while only requiring the experimental realisation of standard qubit strategies.

\section{Conclusion}
\label{sec:conclusion}

In this article, we have investigated classical and quantum correlations limited only by the information content of the corresponding classical and quantum communication. This constitutes a departure from conventional dimension bounded communication in favour of an analysis based on entropic quantities. We have presented a complete characterisation of informationally restricted classical correlations in terms of linear programming, thereby generalising the results of \cite{Tavakoli2020} based on deterministic communication models. For the set of informationally restricted quantum correlations, we have both developed efficient interior-point search methods and hierarchies of semidefinite relaxations for placing upper bounds on the set. We have applied these tools to device-independently witness the amount of information carried by a classical and quantum ensemble as well as to establish strict resource inequalities for different information resources. Furthermore, we have outlined a new avenue for semi-device-independent quantum information processing based on the information assumption. This was exemplified through the investigation of semi-device-independent random number generation for which we both reported nearly optimal rates and advantages over dimension bounded systems. The results presented in this work provide important tools for analysing informationally restricted classical and quantum correlations.

Our work leaves a number of open problems, some of which we list here. 1) How tight are the bounds obtained through our semidefinite  hierarchy for informationally restricted quantum correlations? Can one introduce a semidefinite  hierarchy that provably converges to the quantum set?
  2)  Is there a strict resource inequality for informationally restricted classical correlations for the deterministic versus general communication models? 3) It would be interesting to consider the experimental implementation of semi-device-independent random number generation based on the information assumption. 4) Are there other semi-device-independent protocols that are practical to base on the information assumption? Two obvious candidates to consider for this purpose are quantum key distribution and self-testing.			

Finally, we note that the information-restricted approach also can be used as a relaxation method to bound correlations in prepare-and-measure experiments subject to other assumptions, for which methods to bound the set of quantum correlations are not known.

\begin{acknowledgments}
  We thank Nicolas Brunner and Marie Ioannou for discussions. This work was supported by the Swiss National Science Foundation via the NCCR-SwissMap and the EU Quantum Flagship project QRANGE. E.Z.C. and A.T.~acknowledge support by the Swiss National Science Foundation via the Early PostDoc Mobility fellowships P2GEP2 188276 and P2GEP2 194800. S.P. is a Senior Research Associate of the Fonds de la Recherche Scientifique -- FNRS.
\end{acknowledgments}

\bibliography{stochastic}

\appendix

\section{Derivation of vertex probabilities}
\label{sec:vertex-probabilities}

Here we derive the vertex probabilities
\eqref{eq:vertex-v1}--\eqref{eq:vertex-v4} in
Section~\ref{sec:stochastic-same} using the Fourier-Motzkin algorithm. For
convenience, we undertake the derivation for joint probabilities
\begin{IEEEeqnarray}{c+c}
  p_{bx} = q_{x} p(b|x) , & x = 1,2, 
\end{IEEEeqnarray}
in which we absorb the prior probabilities $q_{x}$ with which the inputs $x = 1$ and $x = 2$ are chosen. We also drop Bob's input $y$, since the constraints on these probabilities are just the same repeated for each $y$. These probabilities are characterised by
\begin{IEEEeqnarray}{rCl}
  \sum_{b} p_{bx} &=& q_{x} , \\
  p_{bx} &\geq& 0 , \\
  \sum_{b} \max(p_{b1}, p_{b2}) &\leq& G .
\end{IEEEeqnarray}
Setting $p_{b1} = (v_{b} + \delta_{b})/2$ and $p_{b2} = (v_{b} - \delta_{b})/2$
we can reexpress the same problem as
\begin{IEEEeqnarray}{rCl}
  \label{eq:vb-normalisation}
  1 - \sum_{b} v_{b} &=& 0 , \\
  \delta - \sum_{b} \delta_{b} &=& 0 , \\
  v_{b} + \delta_{b} &\geq& 0 , \\
  v_{b} - \delta_{b} &\geq& 0 , \\
  \label{eq:deltab-system}
  \Delta - \sum_{b} \abs{\delta_{b}} &\geq& 0 ,
\end{IEEEeqnarray}
with $\Delta = 2 G - 1$ and $\delta = q_{1} - q_{2}$. The last inequality
should be read as $2^{n_{B}}$ different linear inequalities, corresponding to
the $2^{n_{B}}$ different combinations of substitutions
$\abs{\delta_{b}} = \pm \delta_{b}$.

The most general valid inequality can be obtained by taking linear
combinations of the above constraints with nonnegative coefficients for the
inequalities and arbitrary coefficients for the equalities, i.e.,
\begin{IEEEeqnarray}{rCl}
  \lambda \Bigro{1 - \sum_{b} v_{b}}
  + \mu \Bigro{\delta - \sum_{b} \delta_{b}}
  && \nonumber \\
  + \sum_{x} \nu_{b1} (v_{b} + \delta_{b})
  + \sum_{x} \nu_{b2} (v_{b} - \delta_{b}) && \nonumber \\
  +\> \sum_{\vect{s} \in \{\pm\}^{\times n_{B}}}
  \xi_{\vect{s}} \Bigro{\Delta - \sum_{b} s_{b} \delta_{b}}
  &\geq& 0
\end{IEEEeqnarray}
subject to the conditions
\begin{IEEEeqnarray}{rCl}
  \nu_{bx} &\geq& 0 , \\
  \xi_{\vect{s}} &\geq& 0 ,
\end{IEEEeqnarray}
where in the last term the summation index
$\vect{s} = (s_{1}, \dotsc, s_{n_{B}})$ is a vector of signs to use in front
of the $\delta_{b}$s. By grouping the constant terms and terms in $v_{b}$ and
$\delta_{b}$ together we can write the most general possible constraint as
\begin{equation}
  \label{eq:vertex-test}
  \gamma + \sum_{b} \alpha_{b} v_{b} + \sum_{b} \beta_{b} \delta_{b} \geq 0
\end{equation}
with
\begin{IEEEeqnarray}{rCl}
  \gamma &=& \lambda + \delta \mu + \Delta \xi , \\
  \alpha_{b} &=& -\lambda + \sigma_{b} , \\
  \beta_{b} &=& -\mu + \varepsilon_{b} - \xi_{b} ,
\end{IEEEeqnarray}
and
\begin{IEEEeqnarray}{rCl}
  \label{eq:sigmax_def}
  \sigma_{x} &=& \nu_{b1} + \nu_{b2} , \\
  \label{eq:epsilonx_def}
  \varepsilon_{x} &=& \nu_{b1} - \nu_{b2} , \\
  \label{eq:beta_def}
  \xi &=& \sum_{\vect{s}} \xi_{\vect{s}} , \\
  \label{eq:betax_def}
  \xi_{b} &=& \sum_{\vect{s}} \xi_{\vect{s}} s_{b} , \\
  \nu_{bx} &\geq& 0 , \\
  \xi_{\vect{s}} &\geq& 0 .
\end{IEEEeqnarray}
From here, our goal is to eliminate variables until we are left with linear
constraints involving only $\gamma$ and the $\alpha_{b}$s and $\beta_{b}$s.
According to \eqref{eq:vertex-test}, we can interpret these as a sufficient
set of points $(v_{b}, \delta_{b})$ to test to determine if
\eqref{eq:vertex-test} is a valid inequality for given values of $\gamma$,
$\alpha_{b}$, and $\beta_{b}$.

We first eliminate the $\nu_{bx}$s. We take the sum and difference of
\eqref{eq:sigmax_def} and \eqref{eq:epsilonx_def} to get
$\sigma_{b} + \varepsilon_{b} = 2 \nu_{b1}$ and
$\sigma_{b} - \varepsilon_{b} = 2 \nu_{b2}$; combined with $\nu_{bx} \geq 0$
this gives the constraints
\begin{equation}
  -\varepsilon_{b} \leq \sigma_{b} \leq \varepsilon_{b}
\end{equation}
directly on $\sigma_{b}$ and $\varepsilon_{b}$ and we can from this point
forget about the $\nu_{bx}$s. Similarly, \eqref{eq:betax_def} is just
expressing that $\xi_{b}$ are the coordinates of a point that is a `convex
combination of the corners $\{\pm 1\}^{\times n_{B}}$ of the
$n_{B}$-dimensional cube, except that the coefficients are normalised to a
number $\sum_{\vect{s}} \xi_{\vect{s}} = \xi$ instead of one. Thus
\eqref{eq:beta_def} and \eqref{eq:betax_def} are equivalent to
\begin{equation}
  - \xi \leq \xi_{b} \leq \xi .
\end{equation}
Our set of inequalities thus simplifies to
\begin{IEEEeqnarray}{rCl}
  \gamma &=& \lambda + \delta \mu + \Delta \xi , \\
  \alpha_{b} &=& -\lambda + \sigma_{b} , \\
  \beta_{b} &=& -\mu + \varepsilon_{b} - \xi_{b} ,
\end{IEEEeqnarray}
subject to
\begin{IEEEeqnarray}{rCl}
  \sigma_{b} - \varepsilon_{b} &\geq& 0 , \\
  \sigma_{b} + \varepsilon_{b} &\geq& 0 , \\
  \xi - \xi_{b} &\geq& 0 , \\
  \xi + \xi_{b} &\geq& 0 .
\end{IEEEeqnarray}
Let us next eliminate $\sigma_{b}$ and $\xi$. We get
\begin{IEEEeqnarray}{rCl}
  c - \lambda - \delta \mu &\geq& \Delta \xi_{b} , \\
  c - \lambda - \delta \mu &\geq& - \Delta \xi_{b} , \\
  \alpha_{b} + \lambda &\geq& \varepsilon_{b} , \\
  \alpha_{b} + \lambda &\geq& -\varepsilon_{b} , \\
  \beta_{b} + \mu &=& \varepsilon_{b} - \xi_{b} .
\end{IEEEeqnarray}
Eliminating $\varepsilon_{b}$ then gives
\begin{IEEEeqnarray}{rCl}
  \gamma - \lambda - \delta \mu - \Delta \xi_{b} &\geq& 0 , \\
  \gamma - \lambda - \delta \mu + \Delta \xi_{b} &\geq& 0 , \\
  \alpha_{b} + \lambda &\geq& 0 \,, \\
  \alpha_{b} - \beta_{b} + \lambda - \mu - \xi_{b} &\geq& 0 , \\
  \alpha_{b} + \beta_{b} + \lambda + \mu + \xi_{b} &\geq& 0 ,
\end{IEEEeqnarray}
and eliminating $\xi_{b}$ gives
\begin{IEEEeqnarray}{rCl}
  \gamma - \lambda - \delta \mu &\geq& 0 , \\
  \label{eq:ablambda_constraint}
  \alpha_{b} + \lambda &\geq& 0 , \\
  \label{eq:+mu_constraint}
  \gamma + \Delta \alpha_{b} + \Delta \beta_{b} - (1 - \Delta) \lambda
  + (\Delta - \delta) \mu
  &\geq& 0 , \\
  \label{eq:-mu_constraint}
  \gamma + \Delta \alpha_{b} - \Delta \beta_{b} - (1 - \Delta) \lambda
  - (\Delta + \delta) \mu
  &\geq& 0 . \IEEEeqnarraynumspace
\end{IEEEeqnarray}
There are at this point only two unwanted variables left, $\lambda$ and
$\mu$. Eliminating $\lambda$ first gives
\begin{IEEEeqnarray}{rCl}
  \gamma + \alpha_{b} - \delta \mu &\geq& 0 , \\
  \gamma + (1 - \Delta) \alpha_{b} + \Delta \alpha_{b'} + \Delta \beta_{b'}
  + (\Delta - \delta) \mu &\geq& 0 , \\  
  \gamma + (1 - \Delta) \alpha_{b} + \Delta \alpha_{b'} - \Delta \beta_{b'}
  - (\Delta + \delta) \mu &\geq& 0 . \IEEEeqnarraynumspace
\end{IEEEeqnarray}
Note that, here, there can be two different values of the output, $b$ and
$b'$, since when we combine \eqref{eq:ablambda_constraint} with
\eqref{eq:+mu_constraint} and \eqref{eq:-mu_constraint} we have to do it for
all possible values of $b$ in both inequalities. Additionally, the first of
the inequalities we are left with above is redundant since it can be derived
by summing the second and third constraints with $b' = b$ and then dividing
by two. Combining the two remaining inequalities to eliminate the last
variable $\mu$ gives the family of inequalities
\begin{IEEEeqnarray}{rCl}
  \label{eq:generators_4bs}
  \gamma + \Bigro{1 + \frac{\delta}{\Delta}} \frac{1 - \Delta}{2} \alpha_{b}
  + \Bigro{1 - \frac{\delta}{\Delta}} \frac{1 - \Delta}{2} \alpha_{b'}
  \nonumber \\
  +\> \Bigro{1 + \frac{\delta}{\Delta}} \frac{\Delta}{2} \alpha_{b''}
  + \Bigro{1 - \frac{\delta}{\Delta}} \frac{\Delta}{2} \alpha_{b'''}
  \nonumber \\
  +\> \Bigro{1 + \frac{\delta}{\Delta}} \frac{\Delta}{2} \beta_{b''}
  - \Bigro{1 - \frac{\delta}{\Delta}} \frac{\Delta}{2} \beta_{b'''}
  &\geq& 0 \IEEEeqnarraynumspace
\end{IEEEeqnarray}
with up to four different indices, $b$, $b'$, $b''$, and $b'''$, but many of
these are redundant. To begin with, we don't need the inequalities with
$b \neq b'$, so the system reduces to
\begin{IEEEeqnarray}{rCl}
  \label{eq:generators_3bs}
  \gamma + (1 - \Delta) \alpha_{b}
  + \frac{\Delta + \delta}{2} \alpha_{b'}
  + \frac{\Delta - \delta}{2} \alpha_{b''}
  \nonumber \\
  +\> \frac{\Delta + \delta}{2} \beta_{b'}
  - \frac{\Delta - \delta}{2} \beta_{b''}
  &\geq& 0 \IEEEeqnarraynumspace
\end{IEEEeqnarray}
since all of the inequalities in \eqref{eq:generators_4bs} can be recovered
by summing two instances of \eqref{eq:generators_3bs} with different values
of $b$ with weights $(1 + \delta/\Delta)/2$ and $(1 - \delta/\Delta)/2$.

Having now eliminated all the unwanted variables we express
\eqref{eq:generators_3bs} as
\begin{equation}
  \label{eq:ineq-vdelta}
  \gamma + \vect{\alpha} \cdot \vect{v} + \vect{\beta} \cdot \vect{\delta}
  \geq 0
\end{equation}
with
\begin{IEEEeqnarray}{rCl+rCl+rCl}
  \label{eq:sufficient-vs}
  v_{b} &=& 1 - \Delta \,, &
  v_{b'} &=& \frac{\Delta + \delta}{2} , &
  v_{b''} &=& \frac{\Delta - \delta}{2} , \\
  \label{eq:sufficient-deltas}
  \delta_{b} &=& 0 , &
  \delta_{b'} &=& \frac{\Delta + \delta}{2} , &
  \delta_{b''} &=& - \frac{\Delta - \delta}{2} \IEEEeqnarraynumspace
\end{IEEEeqnarray}
and all other $v$s and $\delta$s equal to zero. These terms are additive and
combine if some of the indices coincide; for example, if $b = b' \neq b''$
then we use $v_{b} = 1 - \Delta/2 + \delta/2$ and
$\delta_{b} = \Delta/2 + \delta/2$. Eqs.~\eqref{eq:sufficient-vs} and
\eqref{eq:sufficient-deltas} identify give a set of points
$(\vect{v}, \vect{\delta})$ that it is sufficient to test to find if
\eqref{eq:ineq-vdelta}, for some given coefficients $\gamma$,
$\vect{\alpha}$, and $\vect{\beta}$, is a valid inequality for all points
satisfying the system \eqref{eq:vb-normalisation}--\eqref{eq:deltab-system}
of constraints for $\vect{v}$ and $\vect{\delta}$ above. In terms of
$p_{bx} = (v_{b} + \delta_{b})/2$ and $p_{b2} = (v_{b} - \delta_{b})/2$ and
reintroducing $G$ and $q_{1}$ and $q_{2}$ via $\Delta = 2G - 1$,
$\delta = q_{1} - q_{2}$, and $1 = q_{1} + q_{2}$, these correspond to
\begin{IEEEeqnarray}{c+c+c}
  p_{b1} = 1 - G , &
  p_{b'1} = q_{1} + G - 1 , &
  p_{b''1} = 0 \IEEEeqnarraynumspace
\end{IEEEeqnarray}
and
\begin{IEEEeqnarray}{c+c+c}
  p_{b2} = 1 - G , &
  p_{b'2} = 0 , &
  p_{b''2} = q_{2} + G - 1 . \IEEEeqnarraynumspace
\end{IEEEeqnarray}
Considering different ways of taking the outputs $b$, $b'$, and $b''$ the
same as or different from each other gives five different kinds of
probability distributions, up to relabelling the output. In matrix notation
like we used in Section~\ref{sec:stochastic-same} they are
\begin{IEEEeqnarray}{l}
  \begin{pmatrix}
    q_{1} & 0 & 0 \\
    q_{2} & 0 & 0
  \end{pmatrix} \,, \\ \nonumber \\
  \begin{pmatrix}
    q_{1} & 0 & 0 \\
    1 - G & q_{2} + G - 1 & 0
  \end{pmatrix} \,, \\ \nonumber \\
  \begin{pmatrix}
    1 - G & q_{1} + G - 1 & 0 \\
    q_{2} & 0 & 0
  \end{pmatrix} \,, \\ \nonumber \\
  \label{eq:notvertex}
  \begin{pmatrix}
    1 - G & q_{1} + G - 1 & 0 \\
    1 - G & q_{2} + G - 1 & 0
  \end{pmatrix} \,, \\ \nonumber \\
  \begin{pmatrix}
    1 - G & q_{1} + G - 1 & 0 \\
    1 - G & 0 & q_{2} + G - 1
  \end{pmatrix} \,.
\end{IEEEeqnarray}
With the exception of \eqref{eq:notvertex}, which is not a vertex, dividing
the first and second rows by the prior probabilities $q_{1}$ and $q_{2}$
gives the vertices asserted in Section~\ref{sec:stochastic-same}. We can see
that \eqref{eq:notvertex} is not a vertex by noticing that it can be obtained
from some of the other matrices above. Specifically,
\begin{IEEEeqnarray}{rCl}
  \begin{pmatrix}
    1 - G & G - q_{1} \\
    1 - G & G - q_{2}
  \end{pmatrix}
  &=& \theta_{1} \begin{pmatrix}
    0 & q_{1} \\
    0 & q_{2}
  \end{pmatrix}
  + \theta_{2} \begin{pmatrix}
    q_{1} & 0 \\
    1 - G & G - q_{2}
  \end{pmatrix} \nonumber \\
  &&+\> \theta_{3} \begin{pmatrix}
    1 - G & G - q_{1} \\
    q_{2} & 0
  \end{pmatrix}
\end{IEEEeqnarray}
for the convex coefficients
\begin{IEEEeqnarray}{rCl}
  \theta_{1}
  &=& \frac{(G - q_{1}) (G - q_{2})} {q_{1} q_{2} - (1 - G)^{2}} , \\
  \theta_{2} &=& \frac{(1 - G) (G - q_{1})}{q_{1} q_{2} - (1 - G)^{2}} , \\
  \theta_{3} &=& \frac{(1 - G) (G - q_{2})}{q_{1} q_{2} - (1 - G)^{2}} .
\end{IEEEeqnarray}

\section{Characterisation of the $(2, 1, 2)$ scenario}
\label{sec:212scenario-supp}

\subsection{Characterisation of $\cdet$}

Let us begin by considering this scenario when there is no shared randomness. In that case, the problem is trivial, Alice's messages depend only on $x$, and if the guessing probability bound $G$ is anything strictly less than one then the only possibility is that Alice sends the same message in both cases, in which case the resulting probabilities must be the same. Therefore without shared randomness, the correlations set collapses to a line $E_{1} = E_{2}$.

In the following we suppose that $q_{1} > q_{2}$ without loss of generality. If shared randomness is available, then Alice can sometimes send the same message (with associated guessing probability $q_{1}$) and sometimes send different messages (with guessing probability one) as long as the average guessing probability remains smaller than $G$.

If Alice sends the same message, Bob can generate the following extremal probabilities
\begin{equation}
  \label{eq:E1E2_samemsg}
  (E_{1}, E_{2}) = (+1, +1) \text{ or } (-1, -1) ,
\end{equation}
while if Alice sends different messages Bob can generate the extremal
probabilities
\begin{equation}
  \label{eq:E1E2_diffmsg}
  (E_{1}, E_{2}) = (+1, +1), (+1, -1), (-1, +1), \text{ or } (-1, -1) .
\end{equation}
The extremal probabilities that Bob can generate overall are combinations of
\eqref{eq:E1E2_samemsg} with some probability $\theta$ and \eqref{eq:E1E2_diffmsg}
with some probability $1 - \theta$. We should use
\begin{IEEEeqnarray}{c+t+c}
  \theta = \frac{1 - G}{q_{2}} &and& 1 - \theta = \frac{G - q_{1}}{q_{2}}
\end{IEEEeqnarray}
which are chosen such that
\begin{equation}
  \theta \cdot q_{1} + (1 - \theta) \cdot 1 = G ,
\end{equation}
in order to respect the guessing probability bound of $G$ on average. (We could make these inequalities rather than equalities, but this is unnecessary since \eqref{eq:E1E2_diffmsg} includes the two extremal points in \eqref{eq:E1E2_samemsg} and any excess in the value of $\theta$ could be absorbed into that.) After eliminating two redundant ones this yields six vertices,
\begin{IEEEeqnarray}{rCl}
  (E_{1}, E_{2}) &=& (+\mu, +1), (+1, +1), (+1, +\mu), (-\mu, -1),
  \nonumber \\
  && (-1, -1), (-1, -\mu)
\end{IEEEeqnarray}
with
\begin{equation}
  \mu = \frac{1 + q_{1} - 2 G}{q_{2}} .
\end{equation}
All probabilities represented by values $(E_{1}, E_{2})$ in this scenario
must be convex combinations of these six vertices. In addition to the trivial
conditions $\abs{E_{x}} \leq 1$, this implies two facet inequalities,
\begin{equation}
  \abs{E_{1} - E_{2}} \leq 2 \frac{G - q_{1}}{q_{2}} .
\end{equation}

\subsection{Characterisation of $\qdet$}

The problem is very similar to a quantum set studied in Section~3.1 in \cite{VanHimbeeck2017}. For pure states, the guessing probability associated to the ensemble $\mathcal{E} = \{q_{1}, \psi_{1}; q_{2}, \psi_{2}\}$ is
\begin{equation}
  P_{\guess}(X | \mathcal{E})
  = \frac{1}{2} + \frac{1}{2} \sqrt{
    1 - 4 q_{1} q_{2} \babs{\braket{\psi_{1}}{\psi_{2}}}^{2}} .
\end{equation}
Assuming the guessing probability satisfies $P_{\guess}(X | \mathcal{E}) \leq G$ for some bound $G$ and rearranging for the inner product gives
\begin{equation}
  \label{eq:Gamma_pg_bound}
  \babs{\braket{\psi_{1}}{\psi_{2}}}^{2}
  \geq \frac{G (1 - G)}{q_{1} q_{2}} .
\end{equation}
In the following we derive what this implies for a linear combination
\begin{equation}
  W = c_{1} E_{1} - c_{2} E_{2}
\end{equation}
of correlation terms $E_{x} = \Tr[E \psi_{x}]$ with $-\openone \leq E \leq \openone$. We remark first that the witness $W$ is trivial if the coefficients $c_{1}$ and $c_{2}$ are not of the same sign because the positivity constraints $E_{x} \leq 1$ alone imply
\begin{equation}
  c_{1} E_{1} - c_{2} E_{2} \leq \abs{c_{1}} + \abs{c_{2}} ,
\end{equation}
which is trivially attained with $E_{1} = E_{2} = \pm 1$ when the coefficients are of opposite signs. We thus concentrate on the case that $c_{1}$ and $c_{2}$ are both of the same sign. In the rest of this section we suppose without loss of generality that $c_{1}$ and $c_{2}$ are nonnegative and that $c_{1} + c_{2} = 1$. We also suppose for simplicity that $q_{1} \geq q_{2}$.

Bounding $W$ with $c_{1}$ and $c_{2}$ taken to have the same sign gives
\begin{IEEEeqnarray}{rCl}
  c_{1} E_{1} - c_{2} E_{2}
  &=& \Tr \bigsq{E (c_{1} \psi_{1} - c_{2} \psi_{2})} \nonumber \\
  &\leq& \btrnorm{c_{1} \psi_{1} - c_{2} \psi_{2}} \nonumber \\
  &=& \sqrt{(c_{1} + c_{2})^{2}
    - 4 c_{1} c_{2} \babs{\braket{\psi_{1}}{\psi_{2}}}^{2}} \nonumber \\
  &=& \sqrt{1 - 4 c_{1} c_{2} \babs{\braket{\psi_{1}}{\psi_{2}}}^{2}} ,
\end{IEEEeqnarray}
where we substituted $c_{1} + c_{2} = 1$ in the last line. Combining this
with the bound \eqref{eq:Gamma_pg_bound} on
$\abs{\braket{\psi_{1}}{\psi_{2}}}$ in terms of $G$ gives
\begin{equation}
  \label{eq:Wpurebound}
  c_{1} E_{1} - c_{2} E_{2}
  \leq \sqrt{1 - \frac{4 c_{1} c_{2}}{q_{1} q_{2}} G (1 - G)} .
\end{equation}
The inequality \eqref{eq:Wpurebound} gives a tight upper bound on the witness to the left in terms of the guessing probability assuming Alice sends one of two pure states $\ket{\psi_{x}}$ with probabilities $q_{x}$. To generalise to allow shared randomness we need to take the convex hull of the right side of \eqref{eq:Wpurebound}. Fortunately this is straightforward. The right side of \eqref{eq:Wpurebound} is convex if
\begin{equation}
  \frac{c_{1} c_{2}}{q_{1} q_{2}} \geq 1
\end{equation}
and concave otherwise; this can be determined by computing the second derivative of the family of functions $f_{Q}(x) = \sqrt{1 - 4 Q x (1 - x)}$.

The condition identifying convexity is satisfied under two conditions: if $c_{1} \geq q_{1}$ or if $c_{1} \leq q_{2}$ (remember we are supposing $q_{1} \geq 1/2 \geq q_{2}$). In this case we need to interpolate \eqref{eq:Wpurebound} between the extreme values $G = q_{1}$ and $G = 1$. This gives
\begin{equation}
  c_{1} E_{1} - c_{2} E_{2}
  \leq \frac{1 - G}{q_{2}} \abs{c_{1} - c_{2}}
  + \frac{G - p_{1}}{q_{2}} .
\end{equation}
Supposing $c_{1} \geq q_{1} \geq q_{2} \geq c_{2}$ gives
\begin{equation}
  c_{1} E_{1} - c_{2} E_{2}
  \leq \frac{1}{q_{2}} \Bigro{
    (1 - G) (c_{1} - c_{2}) + G - q_{1}} ,
\end{equation}
which simplifies to
\begin{equation}
  \label{eq:Wredundant1}
  c_{1} E_{1} - c_{2} E_{2}
  \leq \frac{1}{q_{2}} \Bigro{c_{1} - q_{1} + (2 G - 1) c_{2}} .
\end{equation}
Most of this family of inequalities is redundant, since it is implied by the
special case with $c_{x} = q_{x}$,
\begin{equation}
  \label{eq:pE_pure_bound1}
  q_{1} E_{1} - q_{2} E_{2} \leq 2 G - 1 ,
\end{equation}
and the trivial inequality $E_{1} \leq 1$. This can be seen by rewriting
\eqref{eq:Wredundant1} as
\begin{equation}
  \frac{c_{1} - q_{1}}{q_{2}} E_{1}
  + \frac{c_{2}}{q_{2}} \bigro{q_{1} E_{1} - q_{2} E_{2}}
  \leq \frac{c_{1} - q_{1}}{q_{2}} + \frac{c_{2}}{q_{2}} \bigro{2 G - 1} .
\end{equation}
Similarly, if $c_{2} \geq q_{1} \geq q_{2} \geq c_{1}$, we get a family of
inequalities that is the same as \eqref{eq:Wredundant1} except with $c_{1}$
and $c_{2}$ interchanged on the right side,
\begin{equation}
  \label{eq:Wredundant2}
  c_{1} E_{1} - c_{2} E_{2}
  \leq \frac{1}{q_{2}} \Bigro{c_{2} - q_{1} + (2 G - 1) c_{1}} ,
\end{equation}
but only the special case with $c_{1} = q_{2}$ and $c_{2} = q_{1}$, i.e.,
\begin{equation}
  \label{eq:pE_pure_bound2}
  q_{2} E_{1} - q_{1} E_{2} \leq 2 G - 1 ,
\end{equation}
is not implied by other inequalities. This confirms that the only nontrivial linear inequalities satisfied by correlations in the quantum deterministic set are precisely \eqref{eq:Wpurebound} for $q_{2} \leq c_{1}, c_{2} \leq q_{1}$.

Note that part of the boundary of the quantum set coincides with an ellipse, characterised by
\begin{equation}
  (1 - \gamma) (E_{1} + E_{2})^{2} + \gamma (E_{1} - E_{2})^{2}
  = 4 \gamma (1 - \gamma) ,
\end{equation}
for $\gamma = G (1 - G)/(q_{1} q_{2})$.

\section{Finite message dimension in classical scenarios}
\label{sec:finite-message-dim}

Similarly to appendix~\ref{sec:vertex-probabilities}, we work with joint
probabilities
\begin{equation}
  p_{mx} = q_{x} p(m|x)
\end{equation}
with the priors $q_{x}$ absorbed. These as well as allowed values of the
upper bound $G$ on the guessing probability are characterised by
\begin{IEEEeqnarray}{rCl+l}
  q_{x} - \sum_{m} p_{mx} &=& 0 , & \forall m , \\
  p_{mx} &\geq& 0 , & \forall m, x , \\
  G - \sum_{m} p_{m x_{m}} &\geq& 0 , & \forall \vect{x} = (x_{m}) .
\end{IEEEeqnarray}
The most general family of inequalities implied by this is
\begin{IEEEeqnarray}{rCl}
  \sum_{x} \xi_{x} \Bigro{q_{x} - \sum_{m} p_{mx}}
  + \mu_{mx} \, p_{mx} && \nonumber \\
  +\> \sum_{\vect{x}} \lambda_{\vect{x}} \Bigro{G - \sum_{m} p_{m x_{m}}}
  &\geq& 0 ,
\end{IEEEeqnarray}
for any $\xi_{x}$ and any nonnegative $\mu_{mx}$ and
$\lambda_{\vect{x}}$. We can express this as
\begin{equation}
  \gamma + \sum_{mx} \alpha_{mx} p_{mx} + \beta G \geq 0 .
\end{equation}
where
\begin{IEEEeqnarray}{rCl}
  \gamma &=& \sum_{x} \xi_{x} q_{x} , \\
  \alpha_{mx} &=& -\xi_{x} + \mu_{mx} - \lambda_{mx} , \\
  \beta &=& \lambda , \\
  \label{eq:lambda-lambda_x}
  \lambda &=& \sum_{\vect{x}} \lambda_{\vect{x}} , \\
  \label{eq:lambda_mx-lambda_x}
  \lambda_{mx} &=& \sum_{\vect{x}} \lambda_{\vect{x}} \delta_{xx_{m}} , \\
  \label{eq:lambda_x-pos}
  \lambda_{\vect{x}} &\geq& 0 , \\
  \mu_{mx} &\geq& 0 ,
\end{IEEEeqnarray}
where $\delta_{xx'}$ is the Kronecker delta.

We aim to simplify this system to obtain the most straightforward possible
constraints for $\gamma$, $\alpha_{mx}$, and $\beta$. First, to eliminate the
$\lambda_{\vect{x}}$s, note that \eqref{eq:lambda-lambda_x},
\eqref{eq:lambda_mx-lambda_x}, and \eqref{eq:lambda_x-pos} imply
\begin{IEEEeqnarray}{c+t+c}
  \lambda_{mx} \geq 0 &and& \sum_{x} \lambda_{mx} = \lambda .
  \IEEEeqnarraynumspace
\end{IEEEeqnarray}
Conversely, any variables $\lambda_{mx}$ that satisfy these conditions can be
written in the form \eqref{eq:lambda_mx-lambda_x} with nonnegative
$\lambda_{\vect{x}}$s, for example with
\begin{equation}
  \lambda_{\vect{x}} = \lambda^{-(n_{M}-1)} \prod_{m} \lambda_{m x_{m}} ,
\end{equation}
where $n_{M}$ is the number of different values of the variable
$m$. Substituting also $\lambda = \beta$ simplifies the system to
\begin{IEEEeqnarray}{rCl}
  \gamma &=& \sum_{x} \xi_{x} q_{x} , \\
  \alpha_{mx} &=& -\xi_{x} + \mu_{mx} - \lambda_{mx} , \\
  \beta &=& \sum_{x} \lambda_{mx} , \\
  \lambda_{mx} &\geq& 0 , \\
  \mu_{mx} &\geq& 0 .
\end{IEEEeqnarray}
Eliminating the $\lambda_{mx}$s gives
\begin{IEEEeqnarray}{rCl}
  \gamma &=& \sum_{x} \xi_{x} q_{x} , \\
  \beta &=& \sum_{x} \bigro{-\alpha_{mx} - \xi_{x} + \mu_{mx}} , \\
  \mu_{mx} &\geq& \alpha_{mx} + \xi_{x} , \\
  \mu_{mx} &\geq& 0 ,
\end{IEEEeqnarray}
and eliminating the $\mu_{mx}$s gives
\begin{IEEEeqnarray}{rCl+c}
  \gamma - \sum_{x} q_{x} \xi_{x} &=& 0 , & \\
  \label{eq:b-amx-gamma}
  \beta + \sum_{x} u_{x} \bigro{\alpha_{mx} + \xi_{x}} &\geq& 0 , &
  \forall m, u_{x} \in \{0, 1\} . \IEEEeqnarraynumspace
\end{IEEEeqnarray}

The last step would consist of eliminating the $\xi_{x}$s. Note first that
the instance $u_{x} = 0$ for all $x$ of \eqref{eq:b-amx-gamma} gives an
inequality
\begin{equation}
  \label{eq:beta-geq-zero}
  \beta \geq 0
\end{equation}
which does not involve any of the variables $\xi_{x}$. This corresponds to
the (unique) conic generator
\begin{equation}
  \label{eq:M+-conic-generator}
  \bigro{p(m|x), G} = (0, 1)
\end{equation}
of the polyhedron $\mathcal{M}^{+}$ which, in turn, just expresses a property
of $\mathcal{M}^{+}$ that was already evident from its
definition: we can increase the guessing probability bound component $G$ of
any point $\bigro{p(m|x), G}$ in $\mathcal{M}^{+}$, by adding any nonnegative
multiple of \eqref{eq:M+-conic-generator} to it, and the resulting point will
still be in $\mathcal{M}^{+}$.

For the remaining instances of \eqref{eq:b-amx-gamma}, we seek to bound the
maximum number of different values of the message index $m$ that can appear
in any inequality in the process of eliminating the $n_{X}$ variables
$\xi_{x}$. We rewrite the problem as
\begin{IEEEeqnarray}{rCl}
  \label{eq:c-qx-gamma}
  \gamma &=& \sum_{x} q_{x} \xi_{x} , \\
  \label{eq:gamma-b-amx}
  \sum_{x} u_{x} \xi_{x} &\geq& -\beta - \sum_{x} u_{x} \alpha_{mx} ,
\end{IEEEeqnarray}
to make it clear that the initial inequalities \eqref{eq:gamma-b-amx} all
give lower bounds on the $\xi_{x}$s and the problem can be seen as combining
\eqref{eq:c-qx-gamma} with sums of instances of \eqref{eq:gamma-b-amx} such
that the left side equals $\sum_{x} q_{x} \xi_{x}$. Eliminating first one of
the $\xi_{x}$s, which consists of combining \eqref{eq:c-qx-gamma} with all
the instances of \eqref{eq:gamma-b-amx} in which the chosen variable
$\xi_{x}$ appears with a nonzero coefficient $u_{x}$, yields a system of
inequalities that each involve only one value of $m$. The process of
eliminating the remaining $n_{X} - 1$ variables $\xi_{x}$ can then at worst
double the number of different values of $m$ appearing in the inequalities at
each step. The inequalities we obtain for $\gamma$, $\alpha_{mx}$, and
$\beta$ at the end of this process can thus not involve more than
$2^{n_{X} - 1}$ different values of the index $m$. With the the exception of
\eqref{eq:beta-geq-zero} we can write all of them in the form
\begin{equation}
  \gamma + \sum_{mx} \alpha_{mx} p_{mx} + \beta G ,
\end{equation}
from which we infer that the vertices of $\mathcal{M}^{+}$ are strategies
$\bigro{p(m|x), G}$ in which no more than $2^{n_{X} - 1}$ different messages
$m$ are used in each strategy, i.e., in a matrix notation
\begin{equation}
  \bigro{p(m|x)} = \begin{pmatrix}
    p(1|1) & p(2|1) & \cdots \\
    p(1|2) & p(2|2) & \cdots \\
    \vdots & \vdots & \ddots
  \end{pmatrix}
\end{equation}
the components $p(m|x)$ of the vertices of $\mathcal{M}^{+}$ all have at most
$2^{n_{X} - 1}$ columns containing nonzero entries.

At this point we remark that we have not restricted the number of messages
$m$ used overall, which is simply whatever number $n_{M}$ of different values
of $m$ we allow to appear in the problem from the beginning, since the
$2^{n_{X} - 1}$ messages used in each vertex will generally be different for
each vertex. Remember, however, that we are not interested in the
communication strategies represented by $\mathcal{M}^{+}$ themselves but the
extremal correlations
\begin{equation}
  \label{eq:pb-xy-extremal}
  p(b|x,y) = \sum_{m} p(m|x) p(b|y,m)
\end{equation}
that can ultimately be generated with them, which also depend on Bob's
extremal responses $p(b|y, m)$, and we can use a symmetry of the setting to
reduce the communication strategies we need to consider. In particular, both
the sets of extremal communication strategies $\bigbr{\bigro{p(m|x), G}}$ and
of Bob's extremal responses $\{p(b|y, m)\}$ are symmetric with respect to
relabellings of the messages, under which \eqref{eq:pb-xy-extremal} is also
invariant. We can hence limit the number of messages $n_{M}$ we need to
consider to $2^{n_{X} - 1}$ for the purpose of generating the extremal points
$\bigro{p(b|x,y), G}$ of $\mathcal{C}^{+}$, as allowing more messages will
only result in more ways of generating the same correlations $p(b|x,y)$
through \eqref{eq:pb-xy-extremal}.

\end{document}